\documentclass[10pt,english,prb,twocolumn,superscriptaddress]{revtex4-1}
\usepackage[T1]{fontenc}
\usepackage[latin9]{inputenc}
\usepackage{pdfcolmk}
\usepackage{float}
\usepackage{amsmath}
\usepackage{amssymb}
\usepackage{graphicx}
\PassOptionsToPackage{normalem}{ulem}
\usepackage{ulem}

\makeatletter

\@ifundefined{textcolor}{}
{%
 \definecolor{BLACK}{gray}{0}
 \definecolor{WHITE}{gray}{1}
 \definecolor{RED}{rgb}{1,0,0}
 \definecolor{GREEN}{rgb}{0,1,0}
 \definecolor{BLUE}{rgb}{0,0,1}
 \definecolor{CYAN}{cmyk}{1,0,0,0}
 \definecolor{MAGENTA}{cmyk}{0,1,0,0}
 \definecolor{YELLOW}{cmyk}{0,0,1,0}
}

\makeatother

\usepackage{babel}

\newcommand{\ud}{\,\mathrm{d}}
\newcommand{\rd}{\mathrm{d}}
\newcommand{\RR}{\mathbb{R}}

\newcommand{\TT}{\mathrm{T}}

\newcommand{\bd}[1]{\boldsymbol{#1}}

\newcommand{\eps}{\epsilon}

\newcommand{\average}[1]{\langle#1\rangle}

\newcommand{\barint}{\kern4pt \raise3.4pt\hbox{\vrule height.6pt
    width7pt} \kern-11pt \int}

\begin{document}

\title{Infinite swapping replica exchange molecular dynamics leads to \\
  a simple simulation patch using mixture potentials}

\author{Jianfeng Lu}

\email{jianfeng@math.duke.edu}

\selectlanguage{english}%

\affiliation{\textsf{\textit{Mathematics Department, Duke
University,}}\\
\textsf{\textit{ Box 90320, Durham, NC 27708-0320}}}

\author{Eric Vanden-Eijnden}

\email{eve2@cims.nyu.edu}

\selectlanguage{english}%

\affiliation{\textsf{\textit{Courant Institute of Mathematical
      Sciences, New York
      University,}}\\
  \textsf{\textit{ 251 Mercer St., New York, NY 10012}}}
\begin{abstract}
  Replica exchange molecular dynamics (REMD) becomes more efficient as
  the frequency of swap between the temperatures is
  increased. Recently in [Plattner \textit{et al.}
  J. Chem. Phys. \textbf{135}, 134111 (2011)] a method was proposed to
  implement infinite swapping REMD in practice. Here we introduce a
  natural modification of this method that involves molecular dynamics
  simulations over a mixture potential. This modification is both
  simple to implement in practice and provides a better, energy based
  understanding of how to choose the temperatures in REMD to optimize
  efficiency. It also opens the door to generalizations of REMD in
  which the swaps involve other parameters than the temperature.
\end{abstract}


\maketitle

\section{Introduction}
\label{sec:intro}

Replica exchange molecular dynamics (REMD) is one of the most popular
methods to accelerate the conformational sampling of large
biomolecules and other complex molecular systems \cite{Hansmann:97,
  SugitaOkamoto, SugitaKitaoOkamoto, Caflisch, EarlDeem}. It can be
viewed as a generalization to molecular dynamics (MD) simulations of
the replica exchange Monte Carlo method \cite{SwendsenWang, Geyer,
  GeyerThompson, HukushimaNemoto, MarinariParisi, Whittington}. The
basic idea of REMD is to evolve concurrently several copies (or
replica) of the system, and periodically swap their temperatures in a
thermodynamically consistent way.  When a replica feels an
artificially high temperature, it explores its conformation space much
faster than it would at the physical temperature; when it feels the
physical temperature, equilibrium averages can be extracted from its
dynamics. In theory, the efficiency of REMD increases when the
frequency of swaps is pushed up to infinity, but reaching this limit
has proven difficult in practice~\cite{Sindhikara,
  AbrahamGready}. Recently in Ref.~\onlinecite{Dupuis}, Plattner
\textit{et al.}  have proposed a way to avoid this difficulty. The
idea is to first establish analytically what the limiting dynamics of
REMD is at infinite swapping frequency, and then implement this
dynamics directly instead of trying to increase the swapping frequency
in the original REMD. Our main purpose here is to show that a natural
reformulation of the technique of Ref.~\onlinecite{Dupuis} leads to a
simple method in which the various replica in infinite swapping REMD
evolve by standard MD over a new potential that is a
temperature-dependent mixture of the original one which couples all the
replica. This new method is simple to implement, and reduces to a
patch of standard MD codes in which several replica of the system are
evolved in parallel using forces that are the original molecular ones
multiplied by factors that involve the energies of the replicas: these
energies are the only quantities that must be communicated between the
replicas as they evolve. The method gives a new perspective on how to
optimally choose the temperatures (including how many of them to
pick), with implications for the original REMD, by analyzing simple
geometrical characteristics of the mixture potential. As we show
below, it also permits to design generalizations of REMD in which
parameters other than the temperature are used to build the mixture
potential.

\section{Infinite swapping REMD}
\label{sec:remd}

For the sake of simplicity we will consider first
the case of a system governed by the overdamped Langevin equation (the
generalization to standard MD will be given in Sec.~\ref{sec:reform} below):
\begin{equation}
  \label{eq:overdamp}
   \dot{\bd{x}} = \bd{f}(\bd{x}) 
  + \sqrt{2  \beta^{-1}}\, \bd{\eta},
\end{equation}
where $\bd{x} \in \RR^{3n}$ denotes the instantaneous position of the
system with $n$ particles, $\bd{f}(\bd{x}) = -\nabla V(\bd{x})$ is the
force associated with the potential $V(\bd{x})$, $\beta= 1/k_B T$ is
the inverse temperature, $\bd{\eta}$ is a $3n$-dimensional white-noise
and we set the friction coefficient to one for simplicity. The
solutions of~\eqref{eq:overdamp} sample the Boltzmann equilibrium
probability density,
\begin{equation}
 \label{eq:eqpdf}
 \rho_\beta(\bd{x}) = Z_\beta^{-1} e^{-\beta V(\bd{x})}
\end{equation}
where $Z_\beta = \int_{\RR^{3n} }e^{-\beta V(\bd{x})}
d\bd{x}$. Assuming we only take two temperatures (the generalization
to more temperatures is considered in Sec.~\ref{sec:ntemp} below), the
idea behind REMD is to replace~\eqref{eq:overdamp} by
\begin{equation}
  \label{eq:overdampmixt}
  \begin{cases}
     \dot{\bd{x}}_1 = \bd{f}(\bd{x}_1) + \sqrt{2
       \beta_1^{-1}(t)}\, \bd{\eta}_1,\\
     \dot{\bd{x}}_2 = \bd{f}(\bd{x}_2) + \sqrt{2
       \beta_2^{-1}(t)}\, \bd{\eta}_2
  \end{cases}
\end{equation}
where $\bd{x}_1(t)$ and $\bd{x}_2(t)$ are the two replica, and
$\beta_1(t)$ and $\beta_2(t)$ are the two temperatures that
alternatively swap between the physical $\beta$ and the artificial
$\bar\beta < \beta$ (so that $k_B \bar T > k_B T$). These swaps are
attempted with frequency~$\nu$, and the ones from $(\beta_1,\beta_2) =
(\beta, \bar \beta)$ to $(\beta_1,\beta_2) = (\bar \beta, \beta)$ are
accepted with probability
\begin{equation}
  \label{eq:2}
  \min\left(\frac{\rho_{\bar \beta}(\bd{x}_1) \rho_{\beta}(\bd{x}_2)}
  {\rho_{\beta}(\bd{x}_1) \rho_{\bar \beta}(\bd{x}_2)},1\right)
\end{equation}
and similarly for the ones from $(\beta_1,\beta_2) = (\bar \beta,
\beta)$ to $(\beta_1,\beta_2) = (\beta, \bar\beta)$. \eqref{eq:2} is
the standard acceptance probability used in Metropolis Monte Carlo
schemes and it guarantees that \eqref{eq:overdampmixt} samples the
following equilibrium probability distribution in
$(\bd{x}_1,\beta_1,\bd{x}_2,\beta_2)$:
  \begin{equation}
    \label{eq:9}
    \begin{aligned}
      &\bd{\varrho}_{\beta,\bar\beta}(\bd{x}_1,\beta_1,\bd{x}_2,\beta_2)\\
      &=
      \rho_{\beta_1}(\bd{x}_1) \rho_{\beta_2} (\bd{x}_2)
      \left(\tfrac12 \delta_{\beta_1,\beta}\delta_{\beta_2,\bar\beta}+
        \tfrac12 \delta_{\beta_1,\bar\beta}
        \delta_{\beta_2,\beta}\right)
    \end{aligned}
  \end{equation}
  where $\delta_{\beta_1,\beta}$ denotes the Kronecker delta function,
  $\delta_{\beta_1,\beta}=1$ if $\beta_1=\beta$ and
  $\delta_{\beta_1,\beta}=0$ otherwise. Summing over
  $(\beta_1,\beta_2)$ then gives the equilibrium density for the
  replica positions alone, which is a symmetrized version of the
  Boltzmann densities at the two temperatures $\beta$ and $\bar
  \beta$:
\begin{equation}
  \label{eq:eqpdfmixt}
  \varrho_{\beta,\bar\beta}(\bd{x}_1,\bd{x}_2) = 
  \tfrac12 \rho_{\beta}(\bd{x}_1) \rho_{\bar \beta} (\bd{x}_2) 
  + \tfrac12 \rho_{\bar\beta}(\bd{x}_1) \rho_{\beta}(\bd{x}_2)
\end{equation}
As a result the ensemble average at the physical temperature of any
observable $A(\bd{x})$ can be estimated from
\begin{widetext}
\begin{equation}
  \label{eq:1eavg}
  \begin{aligned}
    \average{A}_\beta &\equiv \int_{\RR^{3n}} A(\bd{x})
    \rho_\beta(\bd{x}) d\bd{x}\\
    & = \int\limits_{\RR^{3n}\times \RR^{3n} }
    \big(\omega_{\beta,\bar\beta}(\bd{x}_1,\bd{x}_2)  A(\bd{x}_1)
      + \omega_{\bar\beta,\beta}(\bd{x}_1,\bd{x}_2) A(\bd{x}_2)
      \big) \varrho_{\beta,\bar\beta}(\bd{x}_1,\bd{x}_2) 
      d \bd{x}_1d\bd{x}_2\\
      & = \lim_{T\to\infty} \frac1T\int_0^T 
      \big(\omega_{\beta,\bar\beta}(\bd{x}_1(t),\bd{x}_2(t)) A(\bd{x}_1(t))
      + \omega_{\bar\beta,\beta}(\bd{x}_1(t),\bd{x}_2(t)) A(\bd{x}_2(t))
      \big) dt
  \end{aligned}
\end{equation}
\end{widetext}
where we defined the weight
\begin{equation}
  \label{eq:mu}
  \begin{aligned}
    \omega_{\beta,\bar\beta}(\bd{x}_1,\bd{x}_2) &=
    \frac{\rho_\beta(\bd{x}_1) \rho_{\bar \beta}(\bd{x}_2)
    }{\rho_\beta(\bd{x}_1) \rho_{\bar \beta}(\bd{x}_2) +
      \rho_\beta(\bd{x}_2) \rho_{\bar \beta}(\bd{x}_1)}\\
    &=
    \frac{e^{-\beta V(\bd{x}_1) -\bar \beta V(\bd{x}_2)}
    }{e^{-\beta V(\bd{x}_1) -\bar \beta V(\bd{x}_2)} +
      e^{-\bar\beta V(\bd{x}_1) -\beta V(\bd{x}_2)}}
  \end{aligned}
\end{equation}
The estimator~\eqref{eq:1eavg} is slightly different from the one
traditionally used in REMD~\cite{SugitaOkamoto}, but its validity can
be readily checked by inserting~\eqref{eq:eqpdfmixt} and~\eqref{eq:mu}
in~\eqref{eq:1eavg}, and it will prove more convenient for our
purpose.  To quantify the efficiency of this estimator, notice that
\begin{equation}
  \label{eq:2eavg}
  \begin{aligned}
    & \int\limits_{\RR^{3n}\times \RR^{3n} }
    \big(\omega_{\beta,\bar\beta} A(\bd{x}_1)
      + \omega_{\bar\beta,\beta}A(\bd{x}_2)
      \big)^2 \varrho_{\beta,\bar\beta} 
      d \bd{x}_1d\bd{x}_2\\
      &\quad \le 2 \int\limits_{\RR^{3n}\times \RR^{3n} }
    \big(\omega^2_{\beta,\bar\beta}  A^2(\bd{x}_1)
      + \omega^2_{\bar\beta,\beta} A^2(\bd{x}_2)
      \big) \varrho_{\beta,\bar\beta} 
      d \bd{x}_1d\bd{x}_2\\
      &\quad  \le 2 \int\limits_{\RR^{3n}\times \RR^{3n} }
    \big(\omega_{\beta,\bar\beta}  A^2(\bd{x}_1)
      + \omega_{\bar\beta,\beta} A^2(\bd{x}_2)
      \big) \varrho_{\beta,\bar\beta} 
      d \bd{x}_1d\bd{x}_2\\
      & \quad = 4 \average{A^2}_\beta
  \end{aligned}
\end{equation}
where $\omega_{\beta,\bar\beta} \equiv
\omega_{\beta,\bar\beta}(\bd{x}_1,\bd{x}_2) $,
$\omega_{\bar\beta,\beta} \equiv
\omega_{\bar\beta,\beta}(\bd{x}_1,\bd{x}_2) $ and
$\varrho_{\beta,\bar\beta} \equiv
\varrho_{\beta,\bar\beta}(\bd{x}_1,\bd{x}_2)$, and we used the
property that $0\le \omega_{\beta,\bar\beta}\le 1$.  Therefore, the
variance of the estimator~\eqref{eq:1eavg} is controlled by the
variance of $A(\bd{x})$ under the original density
$\rho_\beta(\bd{x})$. This means that the efficiency of this estimator
is determined by how fast the coupled system~\eqref{eq:overdampmixt}
converges towards equilibrium.  As mentioned before, the higher $\nu$,
the faster this convergence is \cite{Sindhikara, AbrahamGready,
  Dupuis, DupuisMMS}, but large values of $\nu$ requires one to make
many swapping attempts, which slows down the simulations. The key
observation made in Ref.~\onlinecite{Dupuis} is that the limit
$\nu\to\infty$ can be taken explicitly. In this limit,
  the fast temperatures are adiabatically slaved to the positions of
  the slow replica, and these replica only feel their average effect.
  This leads to the following closed equation for the replica
  positions replacing~\eqref{eq:overdampmixt}:
\begin{equation}
  \label{eq:overdampmixtlim}
  \begin{cases}
     \dot{\bd{x}}_1 = 
     \bd{f}(\bd{x}_1) + \sqrt{2 
      (\beta^{-1}\omega_{\beta,\bar\beta} 
    + \bar\beta^{-1}\omega_{\bar\beta,\beta})}\, \bd{\eta}_1,\\
     \dot{\bd{x}}_2 = \bd{f}(\bd{x}_2) + \sqrt{2
       (\bar\beta^{-1}\omega_{\beta,\bar\beta}
    + \beta^{-1}\omega_{\bar\beta,\beta})}\, \bd{\eta}_2
  \end{cases}
\end{equation}
Thus, simulating with~\eqref{eq:overdampmixtlim} instead
of~\eqref{eq:overdampmixt} is a concrete way to perform infinite
swapping REMD, and several strategies to perform these simulations
were discussed in Ref.~\onlinecite{Dupuis}. Here we would like to take
advantage of these strategies but simplify their implementation by
modifying~\eqref{eq:overdampmixtlim}. How to do so is explained next.

\section{Reformulation}
\label{sec:reform}

The system~\eqref{eq:overdampmixtlim} is quite complicated to simulate
because it involves a multiplicative noise. Yet because this system
satisfies detailed balance like the original
REMD~\eqref{eq:overdampmixt} does, it has a specific structure that
can be used to simplify it.  To see how, note that the Fokker-Planck
equation for the joint probability density of $\bd{x}_1(t)$ and
$\bd{x}_2(t)$, $\varrho(t,\bd{x}_1,\bd{x}_2)$, can be written as
\begin{equation}
  \label{eq:1}
  \partial_t \varrho(t) = \text{div}  \big( \mathbb{B}\left(
    \varrho(t)\, \text{grad}\, U + k_BT\,\text{grad}\, \varrho(t) \right)\big)
\end{equation}
Here $\text{div}$ and $\text{grad}$ denote, respectively, the
divergence and gradient operators with respect to
$(\bd{x}_1,\bd{x}_2)$, and we defined the mixture potential
\begin{equation}
  \label{eq:mixtpot}
  \begin{aligned}
    & U(\bd{x}_1,\bd{x}_2)\\
    &= -k_BT\ln \left(e^{-\beta V(\bd{x}_1) -\bar
        \beta V(\bd{x}_2)} + e^{-\bar\beta V(\bd{x}_1) -\beta
        V(\bd{x}_2)}\right)
  \end{aligned}
\end{equation}
as well as the tensor $\mathbb{B}\equiv \mathbb{B}(\bd{x}_1,\bd{x}_2)$:
\begin{equation}
  \mathbb{B} = 
  \begin{pmatrix}
   \omega_{\beta,\bar\beta} +
    \beta\bar\beta^{-1} 
    \omega_{\bar\beta,\beta} & 0 \\
    0 & \omega_{\bar\beta,\beta}+\beta\bar\beta^{-1} \omega_{\beta,\bar\beta}  
  \end{pmatrix}.
\end{equation} 
It is easy to check that the stationary solution of~\eqref{eq:1} is
$\varrho(t,\bd{x}_1,\bd{x}_2)=\varrho_{\beta,\bar\beta}(\bd{x}_1,\bd{x}_2)$,
confirming that the limiting equation~\eqref{eq:overdampmixtlim}
samples~\eqref{eq:eqpdfmixt} like \eqref{eq:overdampmixt} does and
therefore can be used in the estimator~\eqref{eq:1eavg}. The
multiplicative nature of the noise in~\eqref{eq:overdampmixtlim} is a
direct consequence of the fact that the tensor $\mathbb{B}$
in~\eqref{eq:1} depends on $\bd{x}_1$ and $\bd{x}_2$. Therefore, a
natural simplification is to replace $\mathbb{B}$ by a constant
tensor, which, for convenience, we will simply take to be the
identity. This substitution does not affect the stationary solution
of~\eqref{eq:1}, but it changes the system of overdamped Langevin
equations this Fokker-Planck equation is associated with. After some
straightforward algebra, it is easy to see that this new system is
\begin{equation}
  \label{eq:overdampmixtlim2}
  \begin{cases}
     \dot{\bd{x}}_1 =  
    (\omega_{\beta,\bar\beta}
    + \beta^{-1}\bar\beta\omega_{\bar\beta,\beta})
    \bd{f}(\bd{x}_1) + \sqrt{2
       \beta^{-1}}\, \bd{\eta}_1,\\
     \dot{\bd{x}}_2 =  (\omega_{\bar\beta,\beta} 
    +
    \beta^{-1}\bar\beta\omega_{\beta,\bar\beta})
    \bd{f}(\bd{x}_2) + \sqrt{2
       \beta^{-1}}\, \bd{\eta}_2.
  \end{cases}
\end{equation}
This system of equations samples~\eqref{eq:eqpdfmixt}
like~\eqref{eq:overdampmixtlim} does, and its solution can be used in
the estimator~\eqref{eq:1eavg}. But in contrast
with~\eqref{eq:overdampmixtlim}, the noise
in~\eqref{eq:overdampmixtlim2} is simply additive like in the original
equation~\eqref{eq:overdamp}. The only things that have changed
in~\eqref{eq:overdampmixtlim2} are the forces, which are the gradients
with respect to $\bd{x}_1$ and $\bd{x}_2$ of the mixture
potential~\eqref{eq:mixtpot}. As can be seen
from~\eqref{eq:overdampmixtlim2}, these gradients involve the original
forces, $\bd{f}(\bd{x}_1)$ and $\bd{f}(\bd{x}_2)$, multiplied by
scalar factors containing the weight~\eqref{eq:mu}. This means that
the only quantities that must be communicated between the replicas are
the potential energies $V(\bd{x}_1)$ and $V(\bd{x}_2)$ that enter this
weight. 

In practice, rather than~\eqref{eq:overdamp} one is typically
interested in systems governed by the Langevin equation
\begin{equation}
  \label{eq:Langevin}
  \begin{cases}
    \dot{\bd{x}} = m^{-1} \bd{p}, \\
    \dot{\bd{p}} = \bd{f}(\bd{x}) - \gamma \bd{p} + \sqrt{2 \gamma m
      \beta^{-1}}\, \bd{\eta},
  \end{cases}
\end{equation}
where $m$ denotes the mass and $\gamma$ the friction coefficient, in
which case the generalization of~\eqref{eq:overdampmixtlim2} reads
\begin{equation}
  \label{eq:Langevinmixtlim2}
  \begin{cases} 
    \dot{\bd{x}}_1 = m^{-1} \bd{p}_1, \\
    \dot{\bd{p}}_1 = (\omega_{\beta,\bar\beta} +
    \beta^{-1}\bar\beta\omega_{\bar\beta,\beta})
    \bd{f}(\bd{x}_1) \\
    \hspace{8em} - \gamma \bd{p}_1 + \sqrt{2 \gamma m
      \beta^{-1}}\, \bd{\eta}_1,\\
    \dot{\bd{x}}_2 = m^{-1} \bd{p}_2, \\
    \dot{\bd{p}}_2 = (\omega_{\bar\beta,\beta} +
    \beta^{-1}\bar\beta\omega_{\beta,\bar\beta}) \bd{f}(\bd{x}_2) \\
    \hspace{8em} - \gamma \bd{p}_2 + \sqrt{2 \gamma m \beta^{-1}}\,
    \bd{\eta}_2,
  \end{cases}
\end{equation}
The solution of these equations can also be used in the
estimator~\eqref{eq:1eavg} and they can be simulated in parallel using
a simple patch of a standard MD code since they too only involve the
modification of the forces discussed above. The extension to molecular
dynamics using other heat baths is straightforward. In the sequel, we
will analyze the performance of~\eqref{eq:Langevinmixtlim2}, test it
on several examples, and generalize this system to situations with
more than two temperatures and where other parameters than the
temperature are used to build the mixture potential.

\section{Efficiency and optimal choice of the temperature~$k_B \bar T$}
\label{sec:efficiency}

\begin{figure}[htb]
  \centering
  \includegraphics[height=4.9cm]{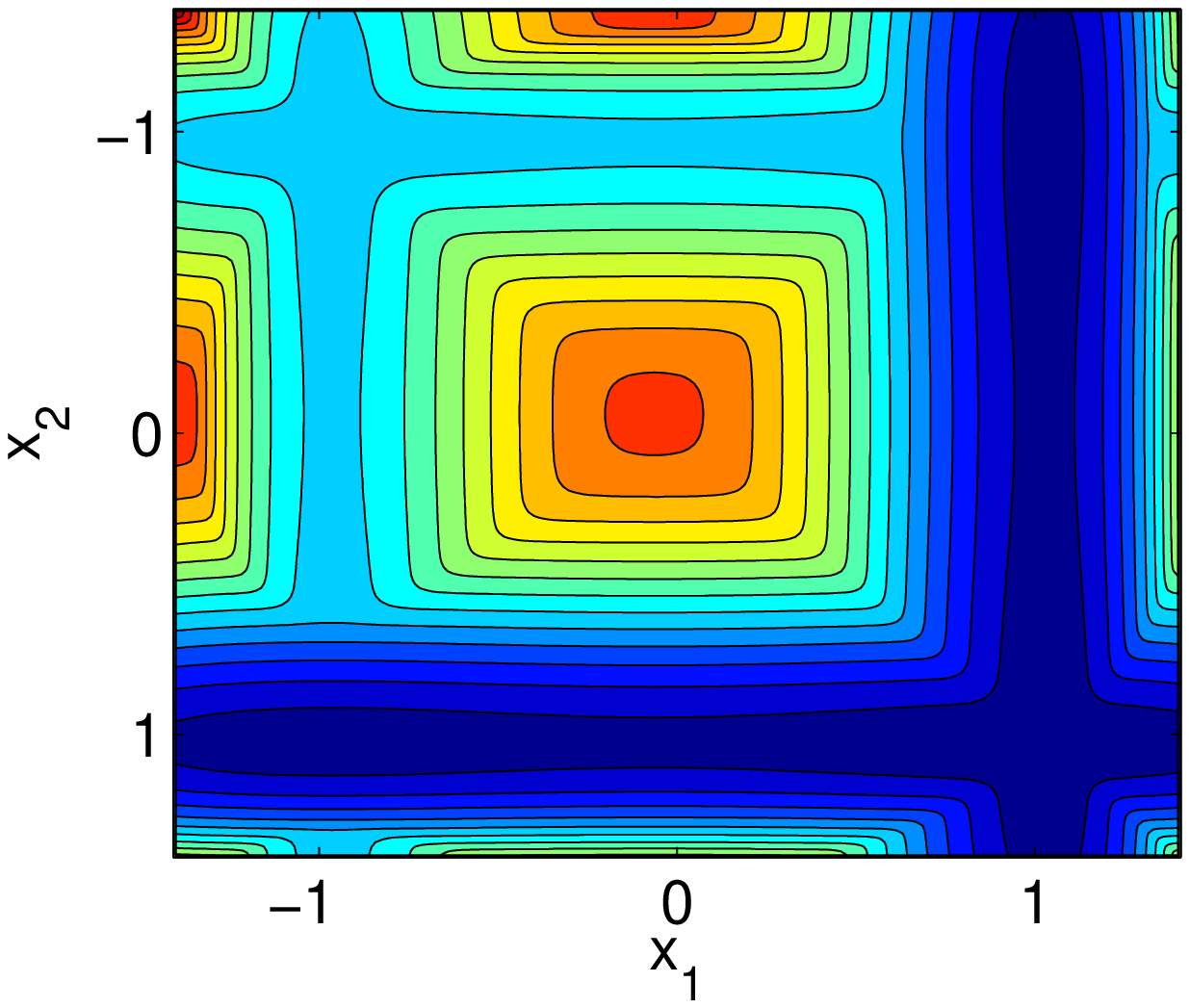}
  \includegraphics[height=4cm]{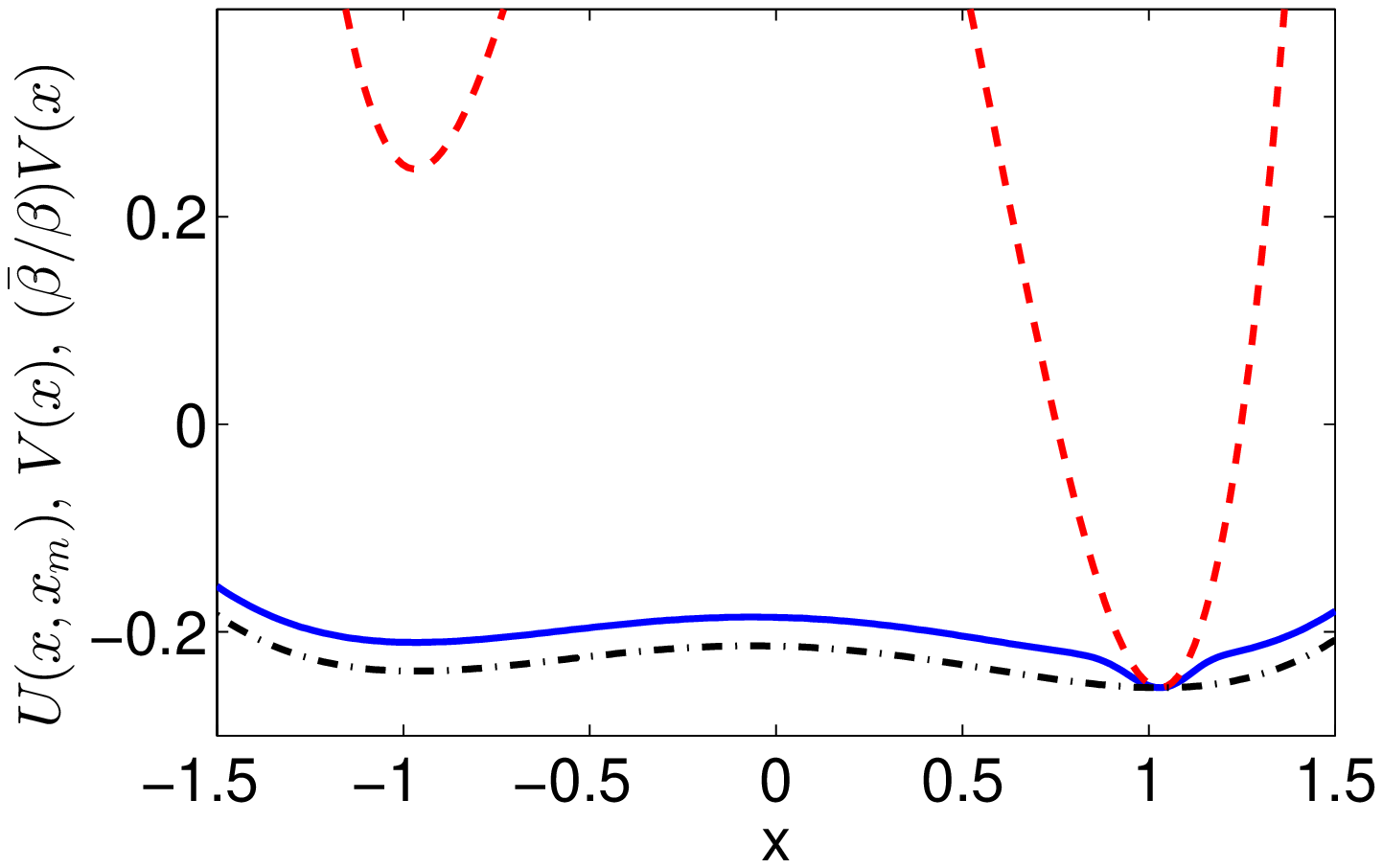}
  \caption{Top panel: The mixture potential \eqref{eq:mixtpot} for the
    potential \eqref{eq:3} clearly showing the two channel (in dark
    blue) connected to the minimum. Bottom panel: A slice of the
    potential along one of the channel (blue solid curve), compared
    with the original potential $V$ (red dashed curve) and its
    scaled-down version $(\bar\beta / \beta) V$ (black dash-dotted
    curve).} \label{fig:1}
\end{figure}

The simple form of \eqref{eq:overdampmixtlim2} or
\eqref{eq:Langevinmixtlim2} allows for a transparent explanation why
these equations are more efficient than the
original~\eqref{eq:overdamp} or~\eqref{eq:Langevin} at sampling the
equilibrium density, and how to choose the artificial temperature $k_B
\bar T = \bar \beta^{-1}$ to optimize this efficiency gain. To see
this, consider a situation in which the original potential $V(\bd{x})$
has a minimum of energy at $\bd{x}_m$. Then the mixture potential
$U(\bd{x}_1,\bd{x}_2)$ has a minimum at $(\bd{x}_m,\bd{x}_m)$, with
two channels connected to it along which this potential is a scaled
down version of the original one, see the top panel of
  Fig.~\ref{fig:1} for an illustration. Indeed, if the artificial
temperature in~\eqref{eq:overdampmixtlim2} is much higher than the
physical one, $\bar \beta \ll \beta$, then in the channel where
$\bd{x}_2\approx\bd{x}_m$ we have $U(\bd{x}_1,\bd{x}_m) \approx
\beta^{-1}\bar \beta V(\bd{x}_1)+ V(\bd{x}_m)$ in the region where
$V(\bd{x}_1)> V(\bd{x}_m)$, and similarly in the channel where
$\bd{x}_1\approx\bd{x}_m$. Thus, along these channels, the equation
for $\bd{x}_1$ in~\eqref{eq:Langevinmixtlim2} can be approximated
locally by
\begin{equation}
  \label{eq:Langevinmixt3}
  \begin{cases}
    \dot{\bd{x}}_1 = m^{-1} \bd{p}_1, \\
    \dot{\bd{p}}_1 = \beta^{-1}\bar \beta
    \bd{f}(\bd{x}_1) - \gamma \bd{p}_1 + \sqrt{2 \gamma m
      \beta^{-1}}\, \bd{\eta}_1,
  \end{cases}
\end{equation}
and similarly for $\bd{x}_2$. \eqref{eq:Langevinmixt3} is like the
original Langevin equation~\eqref{eq:Langevin} except that the force
has been multiplied by a factor $\beta^{-1}\bar\beta \ll1$, meaning
the energy barriers have been lowered by this same factor along the
channels. In essence, by remaining close to a minimum of the energy,
each replica helps the other to surmount barriers and explore the
landscape towards other minima, and this is what accelerates the
sampling. Note however that, while a replica moves fast along a
channel, its weight in the estimator~\eqref{eq:1eavg} is close zero
whereas the one of the replica that hovers near $\bd{x}_m$ is close to
one. Indeed when $\bd{x}_1$ moves and $\bd{x}_2\approx \bd{x}_m$, we
have $\omega_{\beta,\bar\beta}(\bd{x}_1,\bd{x}_m)\approx 0$ and
$\omega_{\bar \beta,\beta}(\bd{x}_1,\bd{x}_m)\approx 1$, and similarly
when $\bd{x}_2$ moves and $\bd{x}_1\approx \bd{x}_m$. Thus we really
need both replica to move in succession, with one of them hovering
near a minimum while the other explores the landscape and vice-versa,
to achieve proper sampling.

Concerning the choice of temperature, the form
of~\eqref{eq:Langevinmixt3} suggests that the optimal $k_B \bar T =
\bar \beta^{-1} $ to pick is the highest energy barrier that the system needs
to surmount to explore its landscape: at lower values of $k_B \bar T$,
crossing this barrier is still a rare event, and at higher values, we
start to blur the sampling by having the system visit regions of too
high energies. As we illustrate next on examples, this intuition is
correct, except that entropic effects also play an important role in
high dimension and may slow down the sampling unless additional
replicas with temperatures between $k_B T$ and $k_B\bar T$ are
introduced (as will be done in Sec.~\ref{sec:ntemp}).

\begin{figure}[htbp]
  \hspace{.3cm}\includegraphics[width=6cm]{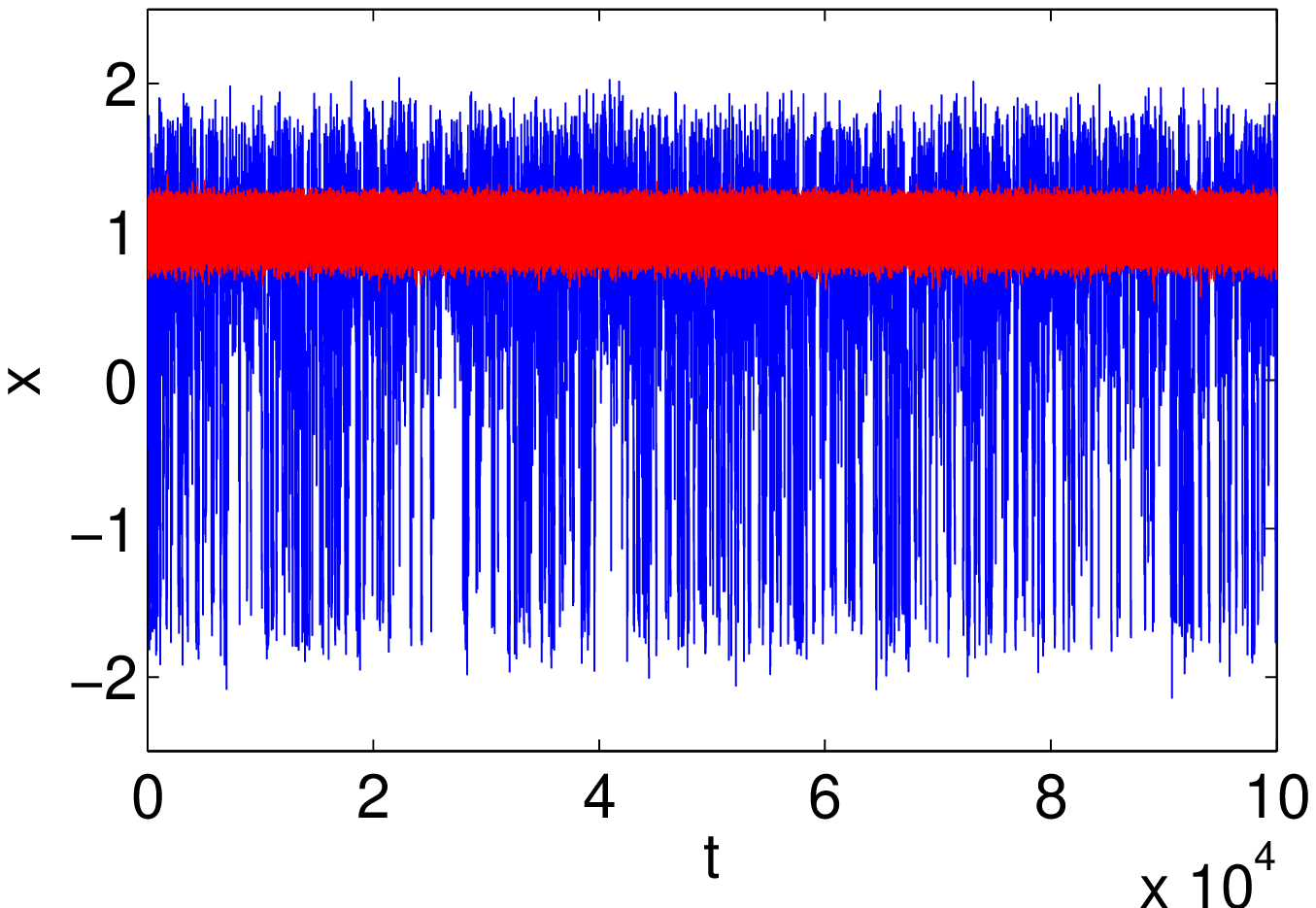}\\
  \includegraphics[width=6cm]{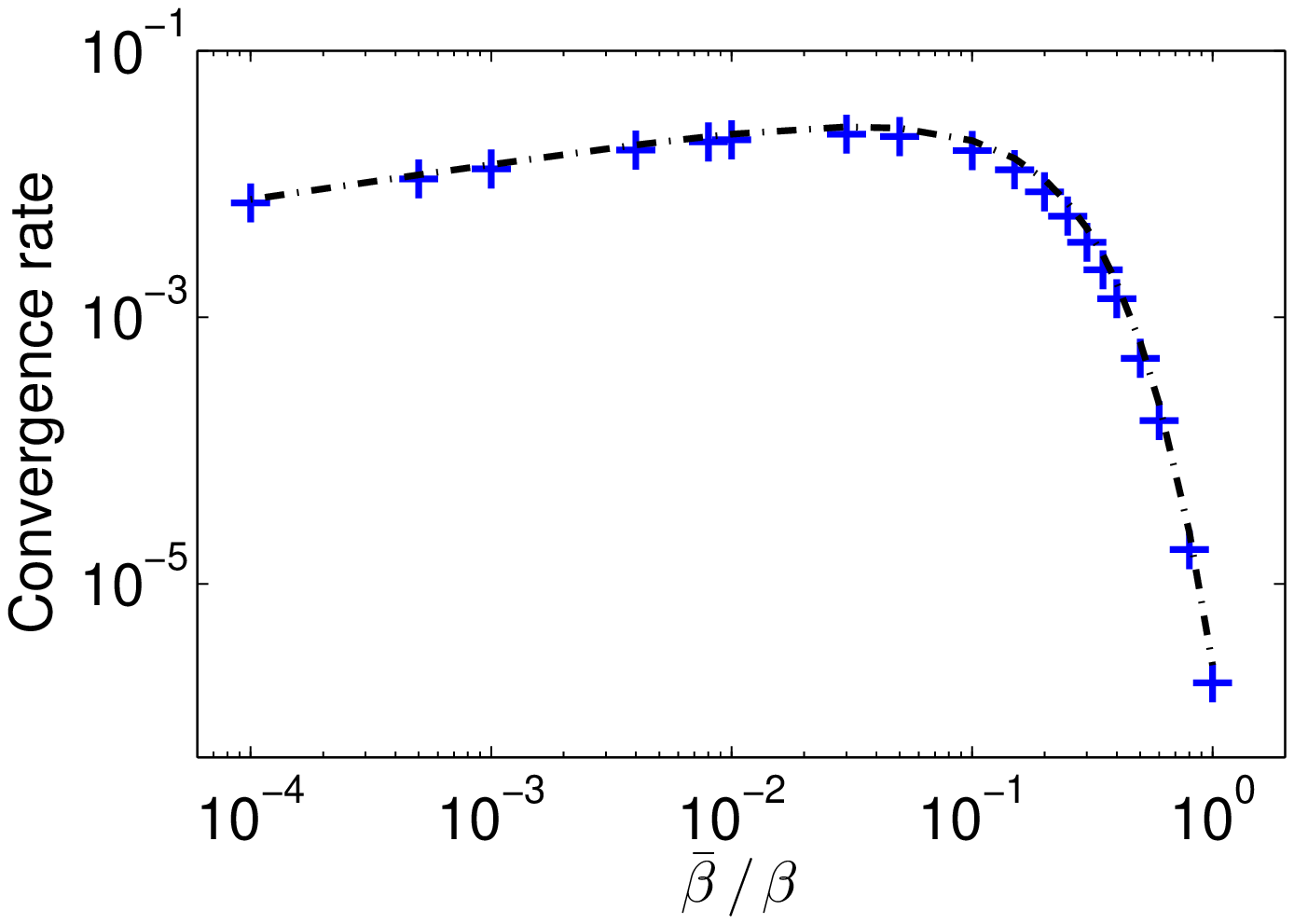} \\
  \includegraphics[width=6.1cm]{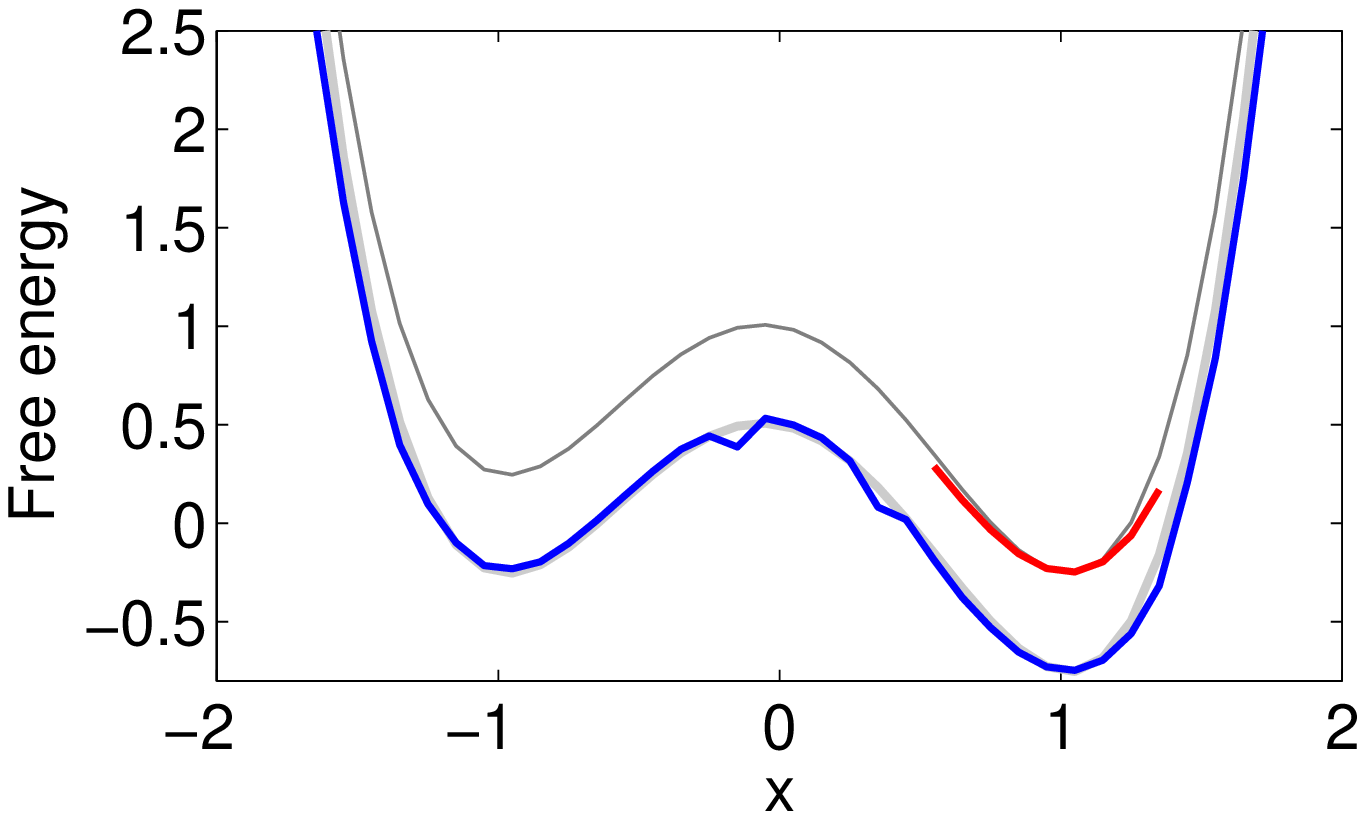}
  \caption{Replica exchange overdamped dynamics for $V(x) = (x^2 -
    1)^2 - \frac14 x$. The physical temperature is $T = \beta^{-1} = 0.04$ and
    the auxiliary high temperature is chosen to be $\bar{T} =
    \bar{\beta}^{-1} = 1.25$, the barrier size. The simulation time is
    $T_{\mathrm{tot}} = 10^5$ with time step $\rd t = 0.025$. 
    Top panel: A typical trajectory (blue) of $x_1(t)$ of the system 
    \eqref{eq:overdampmixtlim2} hops
    between both wells frequently, while a typical trajectory (red)
    under the physical temperature will stay in one of the two wells,
    as the transition is very rare.  
    Middle panel: The convergence rate of the REMD for overdamped dynamics
    \eqref{eq:overdampmixtlim2} with $\beta =
    25$ and different choices of $\bar\beta$.  The blue solid crosses
    show the numerical result, the black dash-dotted curve is the
    estimate obtained from~\eqref{eq:Langevinmixt3} in the high friction limit.
    % , and the red dashed curve is
    % the reaction rate estimated by
    % \eqref{eq:arrhenius1}.
    Bottom panel: The exact free energy (gray solid curves),
    that estimated by \eqref{eq:overdampmixtlim2} (blue solid curve) 
    and that estimated by \eqref{eq:overdamp} (red solid curve,  
    shifted up by $0.1$ to better illustrate the results). 
    \label{fig:2}}
\end{figure}

To test \eqref{eq:overdampmixtlim2} and \eqref{eq:Langevinmixtlim2}
and verify the results above, we first consider a system with
potential
\begin{equation}
  \label{eq:3}
  V(x) = (1-x^2)^2 - \tfrac14 x
\end{equation}
The mixture potential~\eqref{eq:mixtpot} associated with this
$V(\bd{x})$ is plotted in the top panel of
  Fig.~\ref{fig:1}, which clearly shows the two channels mentioned
before. The Bottom panel of Fig.~\ref{fig:1} shows a
slice of the mixture potential along one of the channels and compares
it with $V(x)$ and its scaled-down version $ \beta^{-1}\bar \beta
V(x)$ when $\beta = 25$ (meaning that $k_B T=0.04$ and the energy
barrier to escape the shallow well is about $20 k_BT$ at this physical
temperature) and $\bar \beta = 0.8$. The top panel of
  Fig.~\ref{fig:2} shows the times series of the
original~\eqref{eq:overdamp} and the
modified~\eqref{eq:overdampmixtlim2} for these parameters
values. While the solution of~\eqref{eq:overdamp} is stuck in one
well, that of~\eqref{eq:overdampmixtlim2} explores the two wells
efficiently. The middle panel of Fig.~\ref{fig:2}
shows the convergence rate of~\eqref{eq:overdampmixtlim2} (estimated
from the autocorrelation function of the position) as a function of
$\bar \beta$ and compares it to the analytical estimate of the rate
obtained from~\eqref{eq:Langevinmixt3} in the high friction
limit. This convergence rate reaches a maximum when $\bar \beta =
\Delta V^{-1} \approx 0.8$, consistent with the prediction
from~\eqref{eq:Langevinmixt3}. Finally the bottom
  panel Fig.~\ref{fig:2} shows the free energy reconstructed
using~\eqref{eq:overdampmixtlim2} with $\bar\beta=0.8$ compared to the
one obtained from the original~\eqref{eq:overdamp} with $\beta=25$.

\section{The impact of dimensionality and the need for more than two
  temperatures}
\label{sec:temp}

\begin{figure}[htpb]
  \centering
  \includegraphics[width = 6cm]{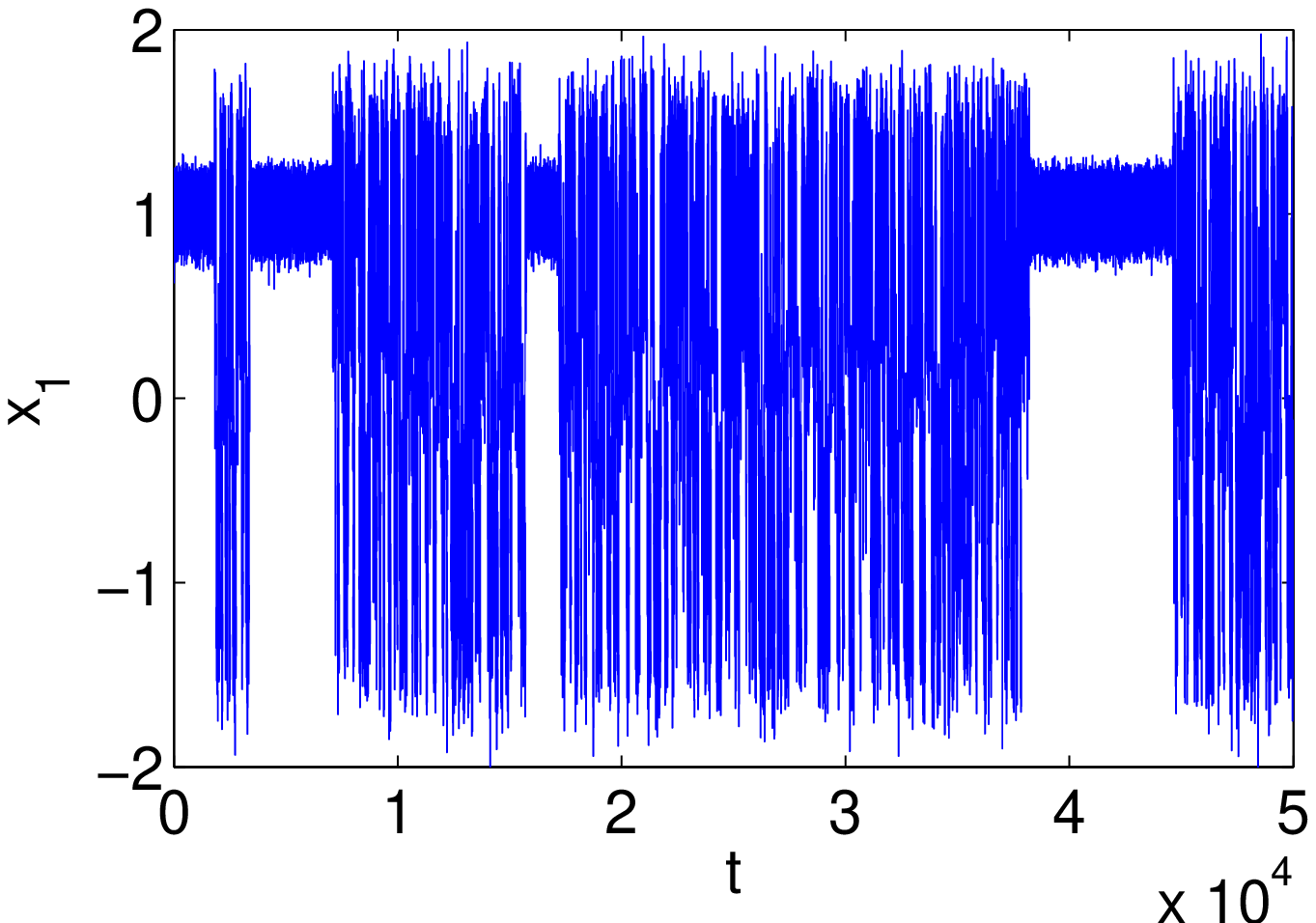}
  \includegraphics[width = 6cm]{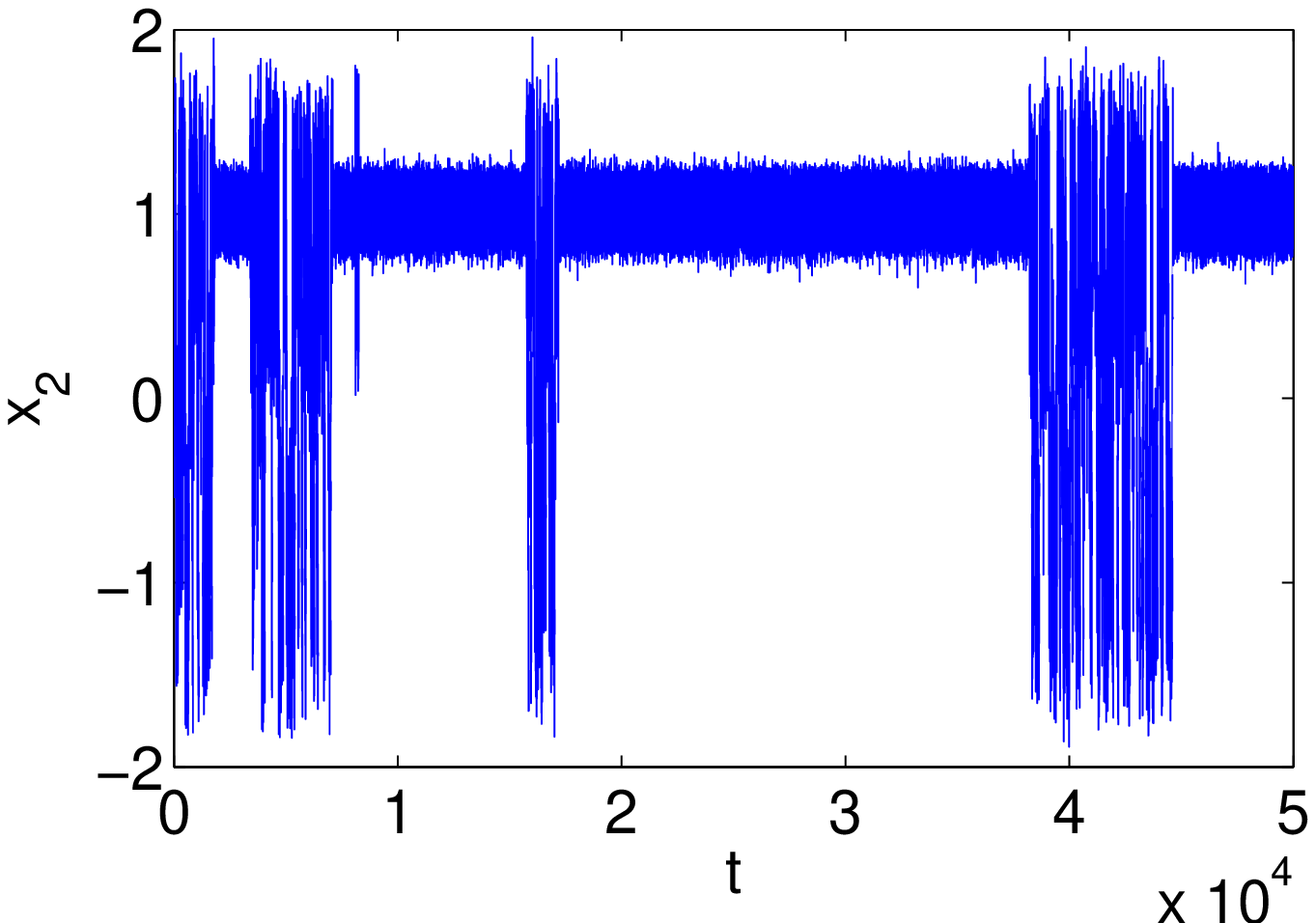}
  \includegraphics[width = 6cm]{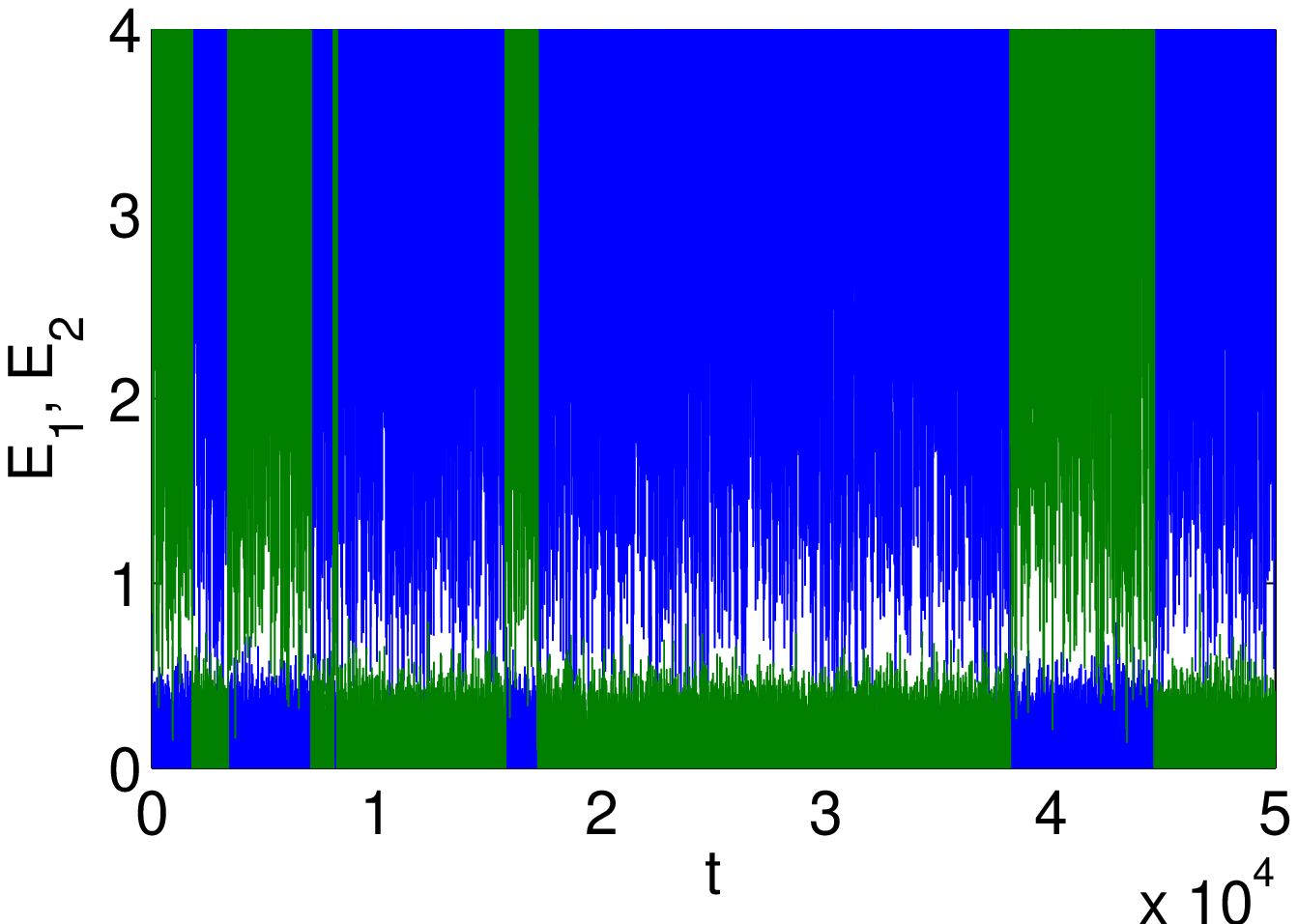}
  \includegraphics[width = 6cm]{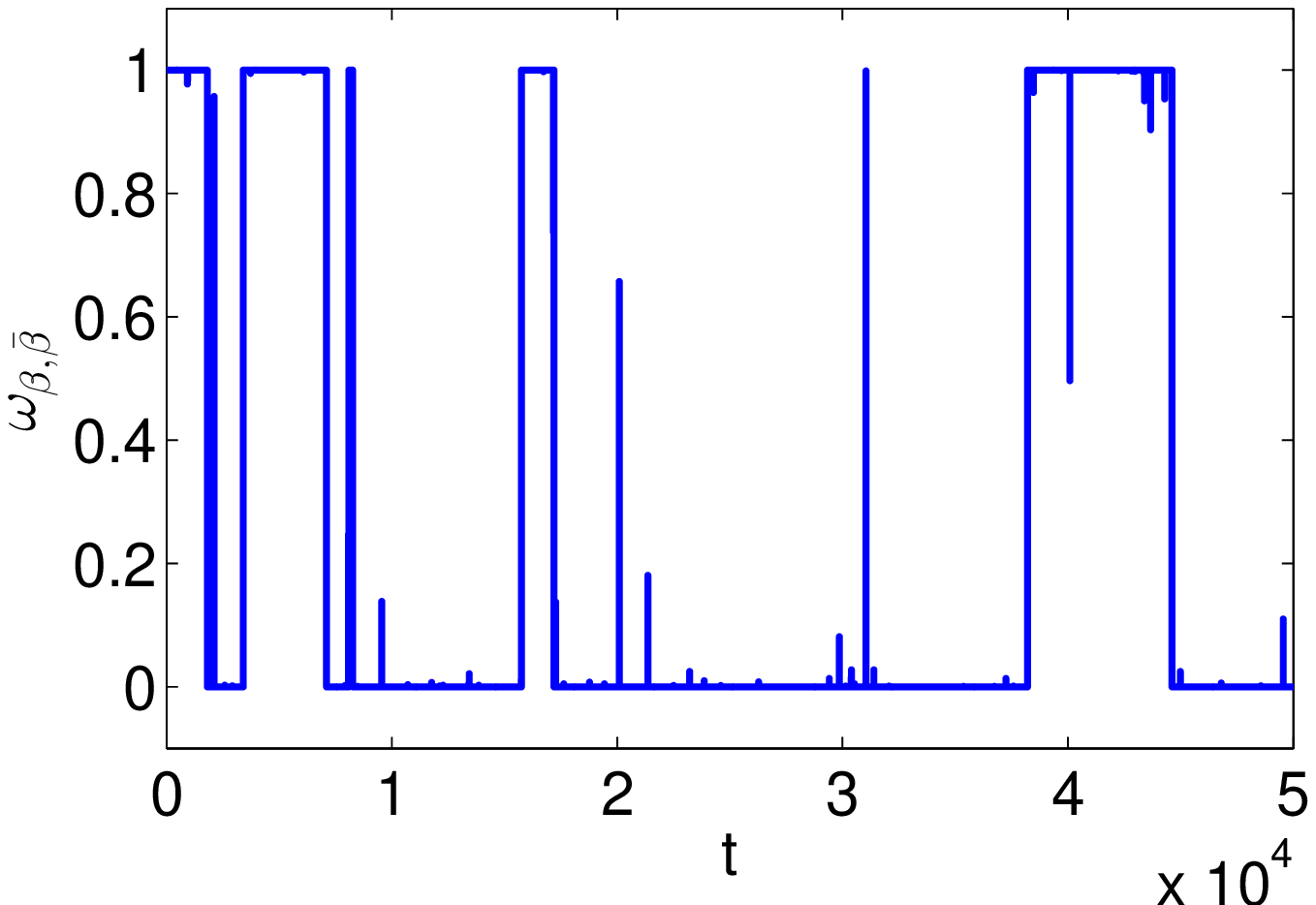}
  \caption{Replica exchange dynamics \eqref{eq:overdampmixtlim2} for
    the potential \eqref{eq:4} with $D = 10$, $\beta = 25$ and
    $\bar\beta = 1$. Top two panels: Typical trajectories of $x_0$ for
    the two replica. Middle panel: Typical trajectories of energies
    for the two replica. Bottom panel: Corresponding weight factor
    $\omega_{\beta, \bar\beta}$ as a function of $t$. The system
    switches between the two channels as $\omega_{\beta, \bar\beta}$
    switches value between $0$ and $1$. This introduces an additional
    slow time scale to the system. \label{fig:highd}}
\end{figure}

\begin{figure}[htpb]
  \includegraphics[width = 6cm]{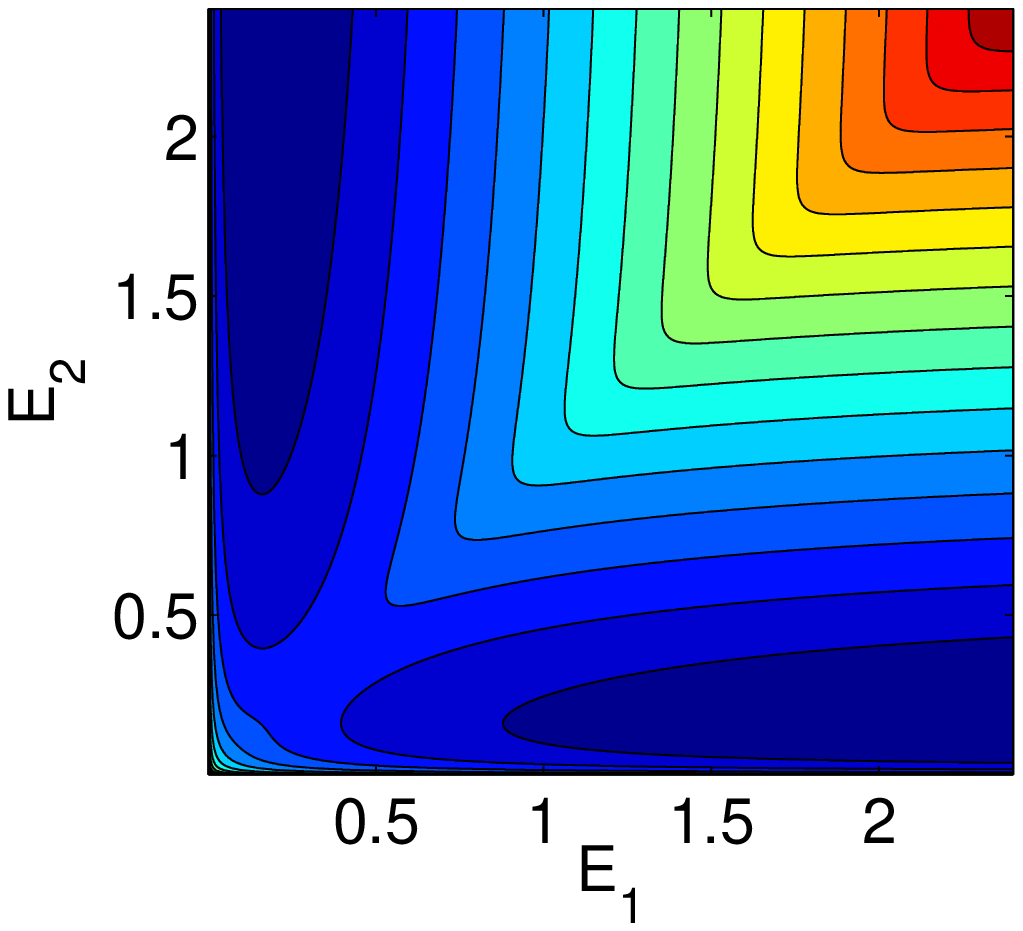}
  \caption{The mixture potential plotted using the energies of the two
    replica as coarse grained variables. The entropic barrier at $E_1
    = E_2$ introduces a slow time scale for switching between
    channels.}
  \label{fig:mixpotE}
\end{figure}

As mentioned in Sec.~\ref{sec:efficiency}, in high dimension entropic
effects start to matter and slow down convergence unless more than two
temperature are used. To analyze the impact of the dimensionality
consider a system with $D$ dimensions moving on the following
potential
\begin{equation}
  \label{eq:4}
  V(x_0, x_1, \ldots, x_{D-1}) = (1-x_0^2)^2 - \frac14 x_0 + 
  \sum_{j=1}^{D-1} \frac12 \lambda_j x_j^2
\end{equation}
where $\lambda_1, \lambda_2, \ldots, \lambda_{D-1}$ are parameters
controlling the curvature of the potential in the $x_1, x_2, \ldots,
x_{D-1}$ directions. In the original equation~\eqref{eq:Langevin}, the
dynamics in the $D$ directions are independent, but this is no longer
the case for the limiting equation~\eqref{eq:Langevinmixtlim2} over
the mixture potential. When the dimensionality is large, $D\gg1$, it
has the effect that the replica moving in the channel
by~\eqref{eq:Langevinmixt3} seldom comes close to a local minimum of
the potential because the basin around this minimum is quite wide; at
the same time, it has to come close enough to one such minimum to
allow the other replica to starts moving in a channel. As can be seen
in Fig.~\ref{fig:highd}, this introduces an additional slow time scale
in the system when $D$ is large, which is related to the presence of an entropic
barrier in the mixture potential. This is shown in
Fig.~\ref{fig:mixpotE} by plotting the free energy $G(E_1,E_2)$ of the
mixture potential $U(\bd{x}_1,\bd{x}_2)$ by using the potential
energies of the two replica as collective variables:
\begin{equation}
  \label{eq:5}
  \begin{aligned}
    G(E_1,E_2) &= - k_B T \ln \int_{\RR^{3n}\times \RR^{3n}} e^{-\beta
      U(\bd{x}_1,\bd{x}_2)} \\
    & \times \delta (V(\bd{x}_1) - E_1) \delta (V(\bd{x}_2) - E_2)
    \ud\bd{x}_1 \ud \bd{x}_2
  \end{aligned}
\end{equation}

\begin{figure}[htpb]
  \centering
  \includegraphics[width=6cm]{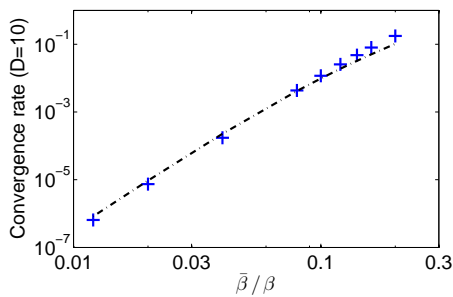}
  \caption{The convergence rate of the REMD for overdamped dynamics
    with $D = 10$ with $\beta = 25$ and different choices of
    $\bar\beta$. The blue solid crosses show the numerical result, the
    black dash-dotted curve is the upper bound \eqref{eq:convrateest}
    obtained from the inverse of mean hitting time
    of~\eqref{eq:Langevinmixt3} to a small ball around the local
    minimum of the potential where the energy is of order $k_BT$ from
    that of this minimum }\label{fig:omegaswitch}
\end{figure}

We can estimate the additional slow time scale to switch from one
channel to the other by calculating the mean time the replica moving
by~\eqref{eq:Langevinmixt3} takes to come within a region near the
local minimum where its potential energy is about $\tfrac{3n}{2} k_BT$
above that of the energy minimum. When this event occurs, the other
replica has a chance to go in his channel and start moving instead,
since $\tfrac{3n}{2} k_BT$ is the typical potential energy of the
system under physical temperature $T$ by equipartition of
energy. However this event becomes less and less likely as the
dimensionality increases because the replica moving
by~\eqref{eq:Langevinmixt3} effectively feels the rescaled potential
$\beta^{-1} \bar \beta V(\bd{x})$ instead of the original one, and so
its potential energy tends to be of order $\tfrac{3n}{2} k_B \bar T$
rather than $\tfrac{3n}{2} k_B T$. Assume that $k_B T = \beta^{-1}$ is
low enough that we can take a quadratic approximation of the potential
near the local minimum, $V(\bd{x}) \approx V(\bd{x}_m) +
\tfrac12(\bd{x}-\bd{x}_m)^{\TT} H (\bd{x}-\bd{x}_m)$. The region that
the moving replica needs to hit is bounded by the ellipsoid defined by
$\tfrac12(\bd{x}-\bd{x}_m)^{\TT} H (\bd{x}-\bd{x}_m)= \tfrac{3n}{2}
k_BT$, and we can use transition state theory to estimate
% \cite{VandenEijndenTal:05}, 
the mean frequency at which the system governed
by~\eqref{eq:Langevinmixt3} hits this ellipsoid:
\begin{equation}\label{eq:convrate}
  \nu = (\det H)^{1/2} (2\pi)^{D/2} \sqrt{2 / \pi\beta} 
  \biggl(\frac{\bar\beta}{\beta}\biggr)^{D/2} 
  e^{-\bar\beta/\beta} \sigma_H, 
\end{equation}
where $D = 3n$ and $\sigma_H$ is the surface area of the ellipsoid
$\tfrac12\bd{x}^{\TT} H \bd{x} = 1$. Using Carlson's bound
\cite{Carlson:66} for $\sigma_H$, we obtain an upper bound
\begin{equation}
\label{eq:convrateest}
  \nu \leq \frac{D^{1/2}(2\pi)^D}{\Gamma((D+1)/2)} 
  \sqrt{\frac{\Lambda}{\pi\beta}} \biggl(\frac{\bar\beta}{\beta}\biggr)^{D/2},
\end{equation}
where $\Lambda$ is the mean curvature of the potential
well. The frequency~$\nu$ also gives the mean rate at
  which the two replica switch from moving fast in the channels or
  remaining trapped near a minimum.  Fig.~\ref{fig:omegaswitch} shows
the convergence rate of~\eqref{eq:overdampmixtlim2} (estimated from
the autocorrelation function of the position) for the
potential~\eqref{eq:4} and shows that this rate is indeed dominated by
the mean hitting frequency in~\eqref{eq:convrateest} when $D$ is
large ($D=10$ for the results reported in the figure: $D=3n$ for
system~\eqref{eq:Langevin}). To avoid this slowing down effect, more
than two temperature must be used, as explained next.

\section{Using multiple temperatures}
\label{sec:ntemp}

The discussion in Sec.~\ref{sec:temp} indicates the need to take more
than two temperatures to accelerate convergence for high dimensional
systems. If we use $N$ temperatures from the physical
  $k_B T$ to the optimal $k_B \bar T$, i.e.
\begin{equation}
  \label{eq:6}
  \beta_1 \equiv \beta =\frac1{k_BT} > \beta_2 > \cdots > \beta_N
  \equiv \bar \beta = \frac1{k_B\bar T},
\end{equation}
then \eqref{eq:mixtpot} generalizes into the following mixture potential
constructed by symmetrization over the $N!$ permutations of the $N$ temperatures
among the $N$ replicas:
\begin{equation}
  \label{eq:8}
  \begin{aligned}
    & U(\bd{x}_1, \ldots, \bd{x}_N) \\
    & = -k_B T \ln \sum_{\sigma} e^{-\beta_1
      V(\bd{x}_{\sigma(1)})\cdots - \beta_N V(\bd{x}_{\sigma(N)})}
  \end{aligned}
\end{equation}
where $\sum_\sigma$ denotes the sum over all the permutations $\sigma$
of the indices $\{1,2,\ldots,N\}$. In turn, the
system~\eqref{eq:Langevinmixtlim2} becomes
 \begin{equation}
   \label{eq:10}
  \begin{cases}
    \dot{\bd{x}}_j = m^{-1} \bd{p}_j, \\
    \dot{\bd{p}}_j = R_j\bd{f}(\bd{x}_j) - \gamma \bd{p}_j + \sqrt{2 \gamma m
      \beta_j^{-1}}\, \bd{\eta}_j,
  \end{cases}
 \end{equation}
where $j=1,\ldots,N$ and $\bd{\eta}_j$ are independent white-noises. Here we defined 
\begin{equation}
  \label{eq:11}
  R_j = \beta^{-1} \sum_\sigma \beta_{\sigma(j)}
  \omega_{\sigma(1),\ldots,\sigma(N)}(\bd{x}_1,\dots,\bd{x}_N) 
\end{equation}
with
\begin{equation}
  \label{eq:12}
  \begin{aligned}
    & \omega_{\sigma(1),\ldots,\sigma(N)}(\bd{x}_1,\dots,\bd{x}_N)\\
    & \quad = \frac{e^{-\beta_1 V(\bd{x}_{\sigma(1)})\cdots - \beta_N
        V(\bd{x}_{\sigma(N)})}} {\sum_{\sigma'} e^{-\beta_1
        V(\bd{x}_{\sigma'(1)})\cdots - \beta_N
        V(\bd{x}_{\sigma'(N)})}}
  \end{aligned}
\end{equation}
If the temperatures in~\eqref{eq:6} are far apart, then at any given
time there typically is one specific permutation $\sigma^*$ such that
$\omega_{\sigma^*(1),\ldots,\sigma^*(N)}(\bd{x}_1,\dots,\bd{x}_N)
\approx 1$ whereas these weights are close to zero for all the other
permutations. This is the multiple replicas equivalent of the slow
switch phenomenon between $\omega_{\beta,\bar \beta}$ and
$\omega_{\bar\beta, \beta}$ being alternatively 1 or 0 that we
observed in Sec.~\ref{sec:temp} with two replicas and it means that
\begin{equation}
  \label{eq:13}
  R_j \approx \beta^{-1} \beta_{\sigma^*(j)}\le1,
\end{equation}
i.e. all the forces in~\eqref{eq:10} are rescaled by factors that are
less or equal to 1. Up to relabeling of the replicas, we can always
assume temporarily that $\sigma^*(j) = j$, meaning that the factors
$R_j$ are ordered as $1=R_1>R_2> \cdots >R_N$. The most likely way for
these factors to change order is that one of the $j$-th replica hits a
small ball where its potential energy becomes of order $k_BT_{j-1}$:
again this is the multiple replica equivalent of the channel switching
process that we observed in Sec.~\ref{sec:temp} with two
replicas. When this process occurs, the permutation $\sigma^*$ for
which the weight is approximately one becomes that in which the
indices $j-1$ and $j$ have been permuted. The frequencies $\nu_j$ at
which these swaps occur can be estimated as in Sec.~\ref{sec:temp}
(compare~\eqref{eq:convrateest}):
\begin{equation}
  \label{eq:convrateestB}
  \nu_j \leq \frac{D^{1/2}(2\pi)^D}{\Gamma((D+1)/2)} 
  \sqrt{\frac{\Lambda}{\pi\beta_{j}}}
  \biggl(\frac{\beta_{j+1}}{\beta_{j}}\biggr)^{D/2}.
\end{equation}
This estimate suggests that we should take a geometric progression of
temperatures in which their successive ratio is kept constant in order
for all the $\nu_j$ (and hence the time scales of channel
switching) to be of the same order:
\begin{equation}
  \label{eq:7}
  \frac{\beta_{j+1}}{\beta_j} =
  \left(\frac{\bar\beta}{\beta}\right)^{1/(N-1)} 
  \qquad j =1, \ldots, N-1 
\end{equation}
This choice agrees with the conventional choice in the literature (see
e.g.~discussions in Refs.~[\onlinecite{Kofke, RathoreChopra,
  RostaHummer, Tavan}]) but gives a different perspective on it.

The discussion above also indicates how many replicas should be
used. Specifically, one should aim at eliminating the slow time scale
of channel switching by taking the successive temperature sufficiently
close together: clearly, in~\eqref{eq:7} the higher $N$, the closer to
1 the ratio $\beta_{j+1}/\beta_j $ becomes even if $\bar\beta/\beta$
is very small. However, this may require taking many replicas, which
in practice poses a difficulty for our approach because the number
$N!$ of terms involved in the weight~\eqref{eq:12} grows very fast
with $N$.

Several strategies can be used to alleviate this problem. For example,
one can decrease the effective dimensionality of the system by only
raising the temperature of a few important degrees of freedom in the
system. This idea was implemented e.g. in Ref.~\onlinecite{Berne} for
biomolecular simulations in solvent.

Another strategy, originally proposed in Ref.~\onlinecite{Dupuis} is
to perform partial swapping. Instead of symmetrizing the potential
over the whole set of the $N$ replicas associated with the $N$
temperatures, the idea is to divide them into several groups
consisting a moderate number of replicas. In each group, a mixture
potential like \eqref{eq:8} with $N$ replaced by the number of replica
in the group is used to evolve the system. To fully mix the
temperatures, multiple partitions are used to distribute the replica
in the different groups and the temperatures in the groups of each
partition are reassigned dynamically.  While none of partition by
itself will fully mix every replica, combining these partitions
together permits to achieve a full mixture of the $N$ replica.

To simplify the presentation of the algorithm, let us consider the
case $N = 3$, with the two partitions given by $A = (12)(3)$ and $B =
(1)(23)$; the extension to the general case is straightforward. We
denote $\alpha_i(t)$ the temperature assigned to the $i$-th replica,
which takes value in $\beta_1=\beta, \beta_2$ and $\beta_3= \bar
\beta$. At the start of the simulation, we set $\alpha_i(0) =
\beta_i$. We then evolve the system using the two partitions $A$ and
$B$ alternatively and dynamically reassign the temperatures as we
switch between the two partitions. This is done by repeating the
following procedures which evolve the system from time $t$ to time $t
+ 2\Delta t$:
\begin{enumerate}
\item Evolve the system using partition $A$ from $t$ to $t+\Delta t$:
  The group of replica $1$ and $2$ is evolved using
  \eqref{eq:Langevinmixtlim2} with the mixture potential
  \begin{multline}
    U_{\alpha_1(t), \alpha_2(t)}(\bd{x}_1, \bd{x}_2) = -
    \beta_1^{-1} \ln (e^{-\alpha_1(t) V(\bd{x}_1)- \alpha_2(t) V(\bd{x}_2)} \\
    + e^{-\alpha_2(t) V(\bd{x}_1) - \alpha_1(t) V(\bd{x}_2)}).
  \end{multline}
  As the other group only consists of replica $3$, it is evolved under
  scaled potential $\alpha_3(t) \beta_1^{-1} V(\bd{x}_3)$ which is
  just the mixture potential with only one replica.
\item At time $t + \Delta t$, reassign the temperatures within each
  group in partition $A$: We set 
  \begin{equation*}
    \begin{cases}
      \alpha_1(t+\Delta t) = \alpha_1(t) \\
      \alpha_2(t+\Delta t) = \alpha_2(t)
    \end{cases}
  \end{equation*}
  with probability
  $\omega_{\alpha_1(t), \alpha_2(t)}(\bd{x}_1, \bd{x}_2)$ and 
    \begin{equation*}
    \begin{cases}
      \alpha_1(t+\Delta t) = \alpha_2(t) \\
      \alpha_2(t+\Delta t) = \alpha_1(t)
    \end{cases}
  \end{equation*}
  with probability $\omega_{\alpha_2(t), \alpha_1(t)}(\bd{x}_1,
  \bd{x}_2)=1-\omega_{\alpha_1(t), \alpha_2(t)}(\bd{x}_1,
  \bd{x}_2)$. Hence, one particular assignment is chosen for each
  group from the symmetrization.
\item Repeat the above two steps to evolve partition $B$ from $t +
  \Delta t$ to $t+2\Delta t$. The replica $1$ is evolved under the
  potential $\alpha_1(t + \Delta t) \beta_1^{-1} V(\bd{x}_1)$. The group of
  replica $2$ and $3$ is evolved with the mixture potential
  \begin{equation}
    \begin{aligned}
      U_{\alpha_2(t+\Delta t), \alpha_3(t+\Delta t)}  &(\bd{x}_2,\bd{x}_3) = \\
      - \beta_1^{-1}
      \ln ( &e^{-\alpha_2(t+\Delta t) V(\bd{x}_2)- \alpha_3(t+\Delta t) V(\bd{x}_3)} \\
      + & e^{-\alpha_2(t+\Delta t) V(\bd{x}_3) - \alpha_3(t+\Delta t)
        V(\bd{x}_2)}). 
    \end{aligned}
  \end{equation}
\item At time $t + 2 \Delta t$, reassign the temperatures within each
  group in partition $B$: We set 
  \begin{equation*}
    \begin{cases}
      \alpha_2(t+2\Delta t) = \alpha_2(t+\Delta t) \\
      \alpha_3(t+2\Delta t) = \alpha_3(t+\Delta t)
    \end{cases}
  \end{equation*}
  with probability
  $\omega_{\alpha_2(t+\Delta t), \alpha_3(t+\Delta t)}(\bd{x}_2, \bd{x}_3)$ and
  \begin{equation*}
    \begin{cases}
      \alpha_2(t+2\Delta t) = \alpha_3(t+\Delta t) \\
      \alpha_3(t+2\Delta t) = \alpha_2(t+\Delta t)
    \end{cases}
  \end{equation*}
  with probability $\omega_{\alpha_3(t+\Delta t), \alpha_2(t+\Delta
    t)}(\bd{x}_2, \bd{x}_3) = 1- \omega_{\alpha_2(t+\Delta t),
    \alpha_3(t+\Delta t)}(\bd{x}_2, \bd{x}_3)$.
\end{enumerate}
This partial mixing strategy can be viewed as a generalization of the
usual replica exchange molecular dynamics in which several replica are
grouped together and evolved under a mixture potential. The parameter
$\Delta t$ is analogous to the inverse of swapping attempt frequency
in the conventional replica exchange. Therefore, it is more
advantageous to take a small $\Delta t$ to increase the swapping
frequency. In practice, we can take $\Delta t$ equal to the time-step used
in the MD simulations.

\begin{figure}[htpb]
  \centering
  \includegraphics[width = 6cm]{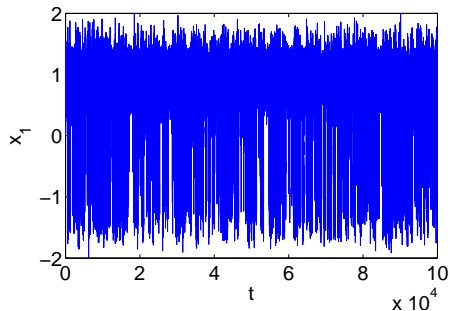}
  \caption{Partial swapping REMD for the potential \eqref{eq:4} with
    $\beta_1 = 25$, $\beta_2 = 5$ and $\beta_3 = 1$. A typical
    trajectory for one of the replica is
    plotted.} \label{fig:partialswapping}

\end{figure}

The performance of the partial swapping algorithm is illustrated in
Fig.~\ref{fig:partialswapping} with three temperatures $\beta_1 = 25$,
$\beta_2 = 5$ and $\beta_3 = 1$. Compared with Fig.~\ref{fig:highd}
the slow time scale of switching between channels is now removed due
to introduction of an intermediate temperature.

\section{Lennard-Jones example}

\begin{figure}[htb]
  \centering
  \includegraphics[width = 6cm]{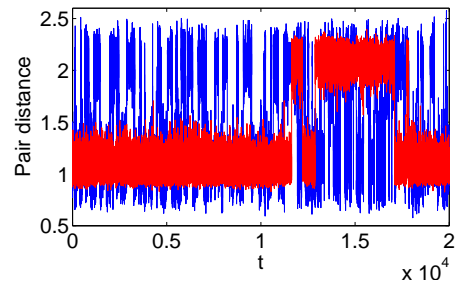}
  \caption{Molecular dynamics example. A typical trajectory of
    the pair distance between the pair of particles interact with
    double well potential in the molecular dynamics simulation under
    physical temperature (red) stays in the compressed and elongated states 
    for a long time, while the trajectory 
    under the dynamics \eqref{eq:Langevinmixtlim2} exhibits frequent 
    transitions between the compressed and elongated states. 
 % (Bottom) The probability density of the pair distance
    % estimated by the coupled dynamics.
%  (blue solid curve) and the molecular 
%     dynamics simulation (red dashed curve). A huge bias is observed in 
%     the molecular dynamics simulation due to the metastability. 
% (Bottom right) 
%     The probability density of the potential energy of the whole system under 
%     temperature $T_1 = 0.2$ (blue solid line) and $T_2 = 1$ (red dashed line).
    \label{fig:MDmix}}
\end{figure}

Finally, to test the performance of our algorithm on a more realistic
example, we apply \eqref{eq:Langevinmixtlim2} to the model system
proposed in Ref.~\onlinecite{DellagoBolhuisChandler}. This system
consists of $N$ two-dimensional particles in a periodic box with side
length~$l$. All particles have the same mass $m$, and they interact
with each other with the Weeks-Chandler-Anderson potential defined as
\begin{equation}
  V_{\text{WCA}}(r) = 
  4 \eps \Bigl( (\sigma / r)^{12} - (\sigma/r)^6 \Bigr) + \eps,
\end{equation}
if $r \leq r_{\text{WCA}} = 2^{1/6} \sigma$, and $V_{\text{WCA}}(r) =
0$ otherwise, except for a pair of particles, which interact via a
double well potential
\begin{equation}
  V_{\text{dW}}(r) = h \Bigl( 1 - \frac{(r - r_{\text{WCA}} - w)^2}{w^2} \Bigr)^2.
\end{equation}
We take $N = 16$, $l = 4.4$, $\sigma = 1$, $h = 1$, $w = 0.5$, and
$\eps = 1$ in the simulation.  The physical temperature is $T = 0.2$,
and for this system, it turns out that it suffices to use two
temperatures with the auxiliary high temperature chosen to be $\bar T
= 1$. The quantity of interest is the free energy associated with the
distance of the pair of particles interacting via the double well
potential. The trajectories of the pair distance is shown in
Figure~\ref{fig:MDmix}, compared to a direct molecular dynamics
simulation. While the original dynamics exhibits metastability in
switching between the compressed and elongated states of the pair
distance, it is observed that the dynamics on the mixture potential
efficiently sample the configurational space. 
% We also plot the potential energy
% distribution of the system under the two temperatures. As in
% conventional replica exchange methods, the efficiency of the method
% depends on the overlap of the distributions.
%

\section{Generalizations}
\label{sec:genral}

\begin{figure}[htb]
  \centering
  \includegraphics[width = 6cm]{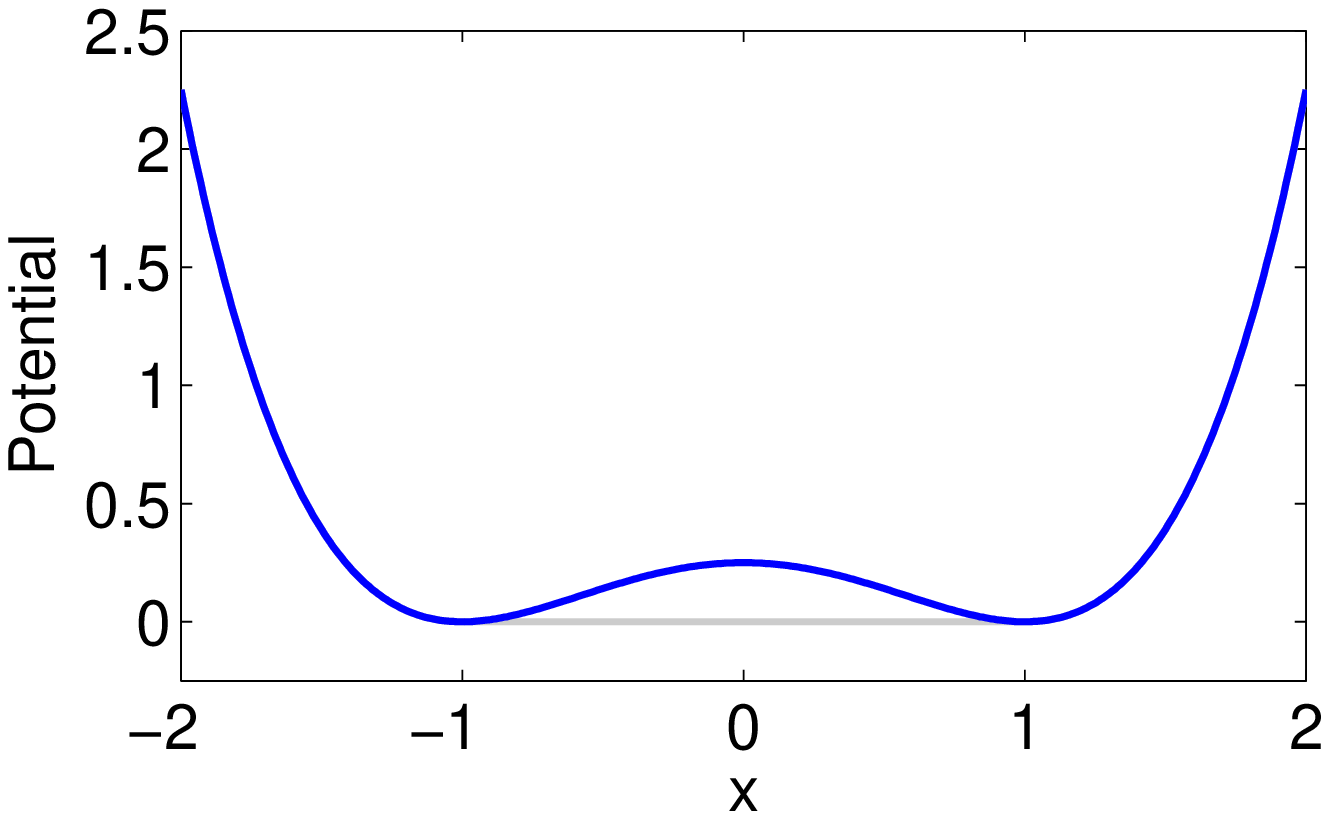} \\
  \includegraphics[width = 6.2cm]{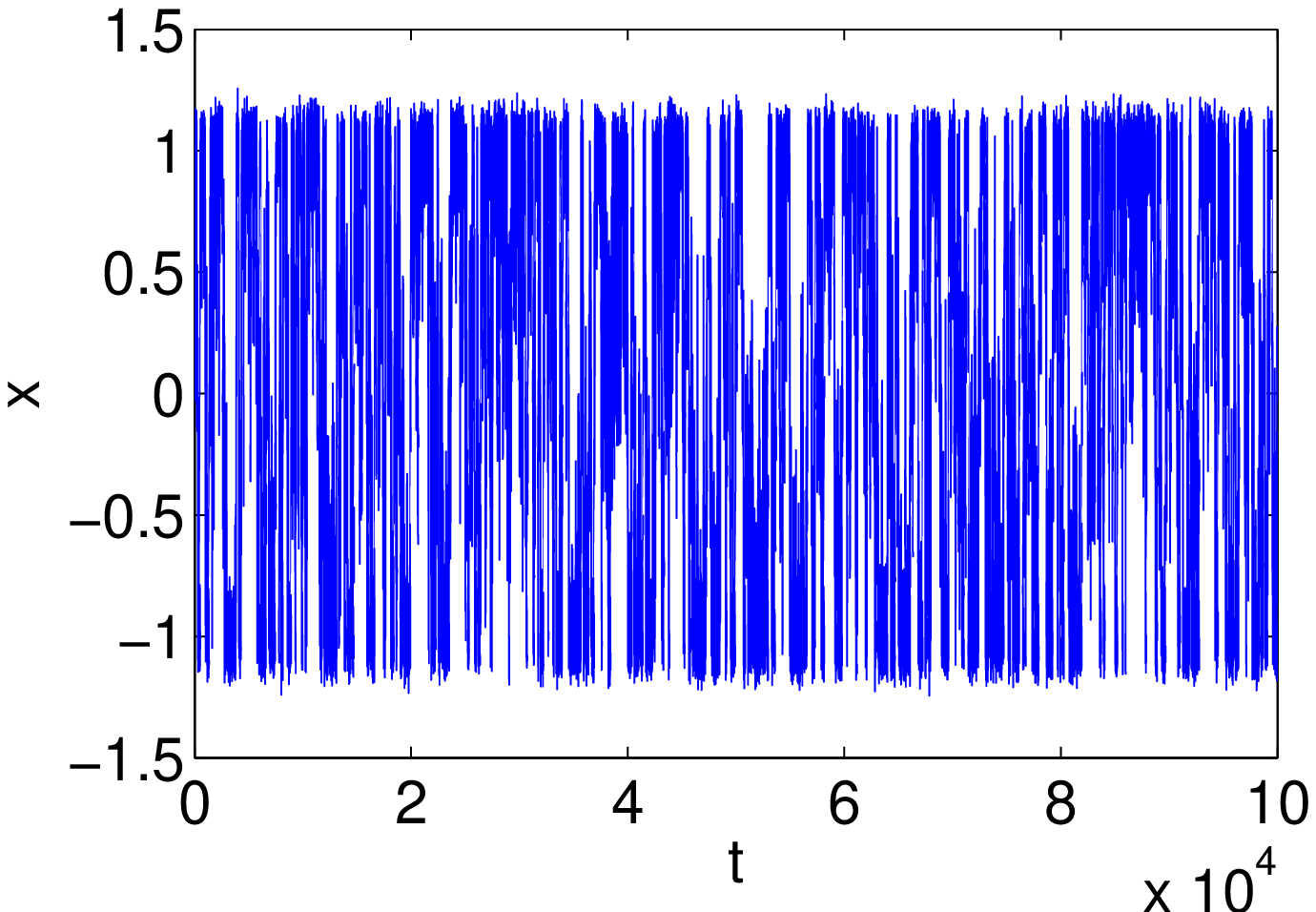}
  \includegraphics[width = 6cm]{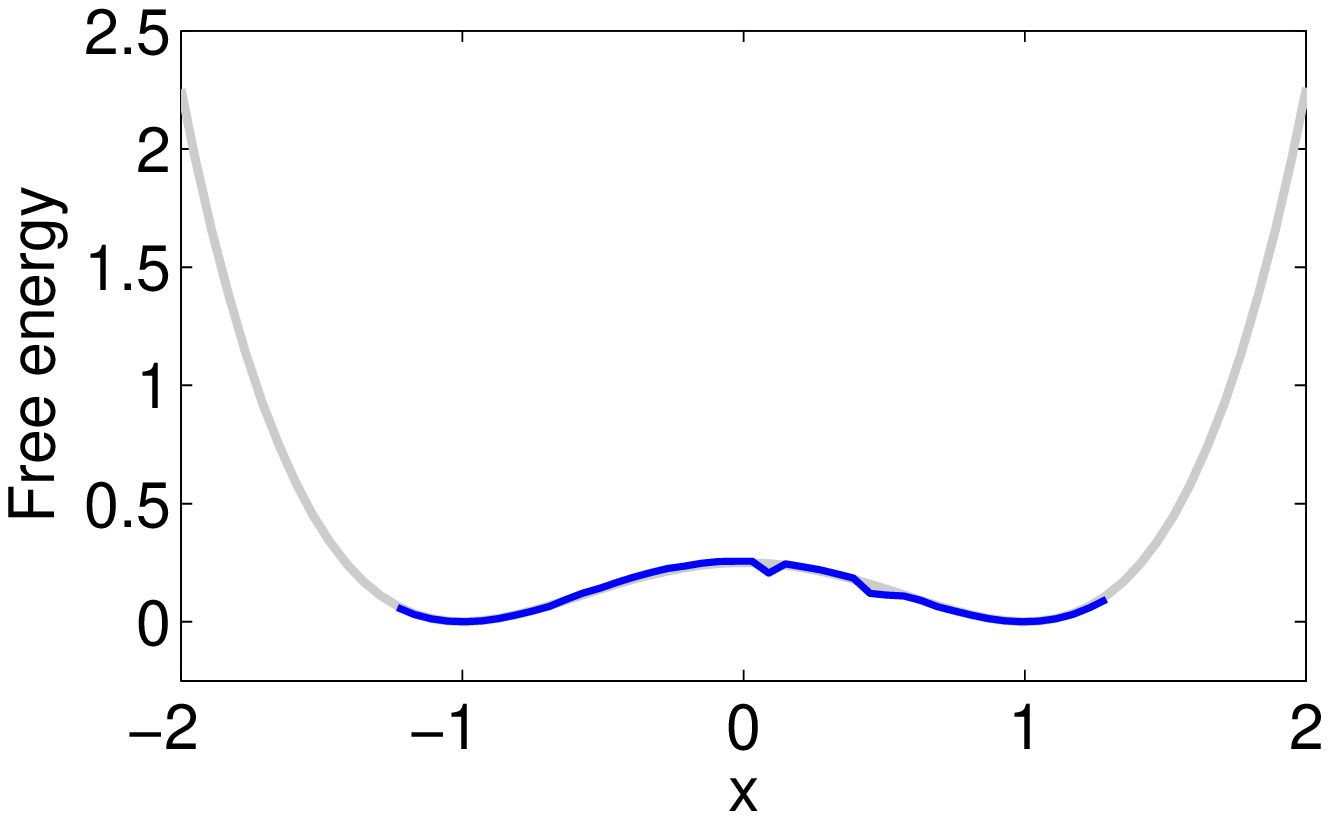}
  \caption{Numerical results for the dynamics \eqref{eq:potchange}.
    Top panel: The physical double well potential $V(x) = (x^2 -
    1)^2/4$ (blue) and the auxiliary potential $\bar{V}$ (gray) where
    the barrier is removed. The other parameters in the model are $m =
    1$, $\gamma = 1$, $\beta = 100$, $T_{\mathrm{tot}} = 1e5$ and $\rd
    t = 0.1$. Middle panel: A typical trajectory of $X_1(t)$ of the
    dynamics \eqref{eq:potchange}.  Since $\bar{V}$ has no barrier,
    the dynamics efficiently explore the region between the two local
    minima of the original potential. Bottom panel: Comparison of the
    estimated (blue) and exact (gray) free
    energies.  \label{fig:potchange}}
\end{figure}

We have used the mixture potential to mix two temperatures in the
above discussion but the idea extends naturally to mixture based on
other parameters. For example, we can mix the original potential with a
modified one in which the barriers between metastable regions is
reduced. Such a modified potential may come from e.g.~spatial warping
\cite{Tuckerman:02} or solute tempering \cite{Berne}. If we denote by
$\bar{V}$ the auxiliary modified potential, the dynamics is given by
\begin{equation}
  \label{eq:potchange}
  \begin{cases}
    \dot{\bd{x}}_1 = m^{-1} \bd{p}_1,  \\
    \dot{\bd{p}}_1 =  (\omega_{V, \bar V} \bd{f}(\bd{x}_1) +
    \omega_{\bar V, V} \bar{\bd{f}}(\bd{x}_1)) \\
    \hspace{8em} - \gamma \bd{p}_1 + \sqrt{2
      \gamma \beta^{-1} m}\ \bd{\eta}_1(t), \\
    \dot{\bd{x}}_2 = m^{-1} \bd{p}_2, \\
    \dot{\bd{p}}_2 =  (\omega_{V, \bar V} \bar{\bd{f}}(\bd{x}_2)
    + \omega_{\bar V, V} \bd{f}(\bd{x}_2)) \\
    \hspace{8em} - \gamma \bd{p}_2 + \sqrt{2 \gamma \beta^{-1} m}\
    \bd{\eta}_2(t),
  \end{cases}
\end{equation}
where $\bd{f}$ and $\bd{\bar{f}}$ are the forces corresponding to the
potentials $V$ and $\bar{V}$ respectively, and the weight
$\omega_{V, \bar V}$ is given by 
\begin{equation}
  \begin{aligned}
    \omega_{V, \bar V}(\bd{x}_1, \bd{x}_2) = \frac{e^{-\beta
        V(\bd{x}_1) - \beta \bar{V}(\bd{x}_2)}}{ e^{-\beta V(\bd{x}_1)
        - \beta \bar{V}(\bd{x}_2)} + e^{-\beta \bar{V}(\bd{x}_1) -
        \beta V(\bd{x}_2)}}
  \end{aligned}
\end{equation}
and similarly for $\omega_{\bar V, V}$.  The
performance~\eqref{eq:potchange} is illustrated on a double-well
example in Figure~\ref{fig:potchange}: here the modified potential
$\bar V$ is simply the original one in which we have removed the
barrier.

\section{Concluding remarks}
\label{sec:conclu}

We have presented a natural reformulation of the infinite swapping
limit of REMD that enables a simple implementation in which forces in
standard MD simulations are rescaled by factors
involving the energies of all the replica. This reformulation is
equivalent to having the system evolve over a mixture potential, and
thereby permits to analyze the efficiency of REMD by using familiar
tools like Arrhenius formula and transition state theory. It also
gives us insights on how to choose an optimal sequence of temperatures
in REMD. Finally, it leads naturally to generalizations in which the
mixture potential is constructed by varying parameters in the
potential other than the temperature, like for example those used in
spatial warping \cite{Tuckerman:02} or solute tempering \cite{Berne}.

\appendix

\section{Dynamics based on Mixture Hamiltonian}

The formulation of the method presented in text involves a mixture
potential, but this mixing can be done on the level of Hamiltonians
too. We briefly describe this alternative in this appendix. Consider a
mixture Hamiltonian of two replica
\begin{equation}
  H(\bd{x}_1, \bd{p}_1, \bd{x}_2, \bd{p}_2) 
  = - k_BT \ln \left( e^{-\beta E_1 -\bar\beta
      E_2} + e^{-\bar\beta E_1
      -\beta E_2} \right).
\end{equation}
where $E_1= E(\bd{x}_1, \bd{p}_1)$ and $E_2=E(\bd{x}_2, \bd{p}_2)$.
The equations of motion associated with this Hamiltonian are 
\begin{equation}
  \label{eq:Hamiltonmix}
  \begin{cases} 
    \dot{\bd{x}}_1 = m^{-1}(\omega_{\beta,\bar\beta} +
    \beta^{-1}\bar\beta\omega_{\bar\beta,\beta}) \bd{p}_1, \\
    \dot{\bd{p}}_1 = (\omega_{\beta,\bar\beta} +
    \beta^{-1}\bar\beta\omega_{\bar\beta,\beta})
    \bd{f}(\bd{x}_1)  \\
    \hspace{8em} - \gamma \bd{p}_1 + \sqrt{2 \gamma m
      \beta^{-1}}\, \bd{\eta}_1,\\
    \dot{\bd{x}}_2 = m^{-1} (\omega_{\bar\beta,\beta} +
    \beta^{-1}\bar\beta\omega_{\beta,\bar\beta}) \bd{p}_2, \\
    \dot{\bd{p}}_2 = (\omega_{\bar\beta,\beta} +
    \beta^{-1}\bar\beta\omega_{\beta,\bar\beta}) \bd{f}(\bd{x}_2) \\
    \hspace{8em} - \gamma \bd{p}_2 + \sqrt{2 \gamma m \beta^{-1}}\,
    \bd{\eta}_2.
  \end{cases}
\end{equation}
where the weight functions $\omega_{\beta, \bar\beta}$ are  given by 
\begin{equation}
  \omega_{\beta,\bar\beta}(\bd{x}_1, \bd{p}_1, \bd{x}_2, \bd{p}_2) 
  = \frac{e^{-\beta E_1 -\bar \beta E_2}
  }{e^{-\beta E_1 -\bar \beta E_2}
    + e^{-\beta E_2 -\bar \beta E_1}}.
\end{equation}
Observe that in the mixture Hamiltonian dynamics, both equations for
$\bd{x}$ and $\bd{p}$ have rescaling terms depending on the weights
$\omega_{\beta, \bar\beta}$.  Thus, compared
with~\eqref{eq:Langevinmixtlim2}, the dynamics
in~\eqref{eq:Hamiltonmix} mixes together both the kinetic and the
potential energies of the two replicas.  As a consequence, in high
dimension, the switching of $\omega_{\beta, \bar\beta}$ from $0$ to
$1$ and vice-versa will be further slowed down. This suggests
that it is more advantageous to use \eqref{eq:Langevinmixtlim2} rather
than \eqref{eq:Hamiltonmix}.

%\FloatBarrier


\end{document}